\documentclass[twocolumn]{aastex63}
\def\actaa{Acta Astronomica}

\usepackage{amsmath}
\hypersetup{linkcolor=red,citecolor=blue,filecolor=green,urlcolor=magenta}

\begin{document}

\shorttitle{SX Phe PL \& PW relations}
\shortauthors{Ngeow et al.}

\title{Zwicky Transient Facility and Globular Clusters: The Period-Luminosity and Period-Wesenheit Relations for SX Phoenicis Variables in the $gri$-Band}

\correspondingauthor{C.-C. Ngeow}
\email{cngeow@astro.ncu.edu.tw}

\author[0000-0001-8771-7554]{Chow-Choong Ngeow}
\affil{Graduate Institute of Astronomy, National Central University, 300 Jhongda Road, 32001 Jhongli, Taiwan}

\author[0000-0001-6147-3360]{Anupam Bhardwaj}
\affil{INAF-Osservatorio astronomico di Capodimonte, Via Moiariello 16, 80131 Napoli, Italy}

\author[0000-0002-3168-0139]{Matthew J. Graham}
\affiliation{Division of Physics, Mathematics, and Astronomy, California Institute of Technology, Pasadena, CA 91125, USA}

\author[0000-0002-7718-7884]{Brian F. Healy} 
\affiliation{School of Physics and Astronomy, University of Minnesota, Minneapolis, MN 55455, USA}

\author[0000-0003-2451-5482]{Russ R. Laher}
\affiliation{IPAC, California Institute of Technology, 1200 E. California Blvd, Pasadena, CA 91125, USA}

\author[0000-0002-0387-370X]{Reed Riddle}
\affiliation{Caltech Optical Observatories, California Institute of Technology, Pasadena, CA 91125, USA} 

\author[0000-0002-9998-6732]{Avery Wold}
\affiliation{IPAC, California Institute of Technology, 1200 E. California Blvd, Pasadena, CA 91125, USA}

\begin{abstract}

  SX Phoenicis (SXP) variables are short period pulsating stars that exhibit a period-luminosity (PL) relation. We derived the $gri$-band PL and extinction-free period-Wesenheit (PW) relations, as well as the period-color (PC) and reddening-free period-Q-index (PQ) relations for 47 SXP stars in located in 21 globular clusters using the optical light curves taken from Zwicky Transient Facility (ZTF). These empirically relations were derived for the first time in the $gri$ filters except for the $g$-band PL relation. We used our $gi$ band PL and PW relations to derive a distance modulus to Crater II dwarf spheroidal which hosts one SXP variable. Assuming that the fundamental and first-overtone pulsation mode for the SXP variable in Crater II, we found distance moduli of $20.03\pm0.23$~mag and $20.37\pm0.24$~mag, respectively, using the PW relation, where the latter is in excellent agreement with independent RR Lyrae based distance to Crater II dwarf galaxy.

\end{abstract}

%\keywords{ {\bf TBD}}

\section{Introduction}\label{sec1}

SX Phoenicis \citep[hereafter SXP; see, for examples,][]{nemec1990,mcnamara1997} variable stars represent a class of low-metallicity and population II pulsating stars with periods in the range of $\sim 0.01$ to $\sim 0.1$~days. These pulsating stars are located in the lower main-sequence on the Hertzsprung-Russell (H-R) diagram, similar to population I $\delta$~Scuti stars -- the counterparts of SXP stars with higher metallicity (near Solar). Sometimes, SXP and $\delta$~Scuti stars are collectively known as dwarf Cepheids. In globular clusters, SXP stars can also occupy the blue stragglers region on the H-R diagram. 

Despite being intrinsically fainter than classical pulsating stars such as Cepheids and RR Lyrae, SXP stars are distance indicators and have been used to derive distance to stellar systems in the past \citep{mcnamara2011}, including Carina dwarf galaxy \citep{mcnamara1995,vivas2013} and a number of globular clusters \citep[for several recent examples, see][]{fj2013,kunder2013,af2015,deras2019,ahumada2021}. This is because SXP stars also obey a period-luminosity (PL) relations similar to other classical pulsating stars. Examples of the empirical PL, or the period-luminosity-metallicity (PLZ), relations for SXP stars can be found in \citet{nemec1990}, \citet{nemec1994}, \citet{mcnamara1995}, \citet{pych2001}, \citet{jeon2003,jeon2004}, \citet{cohen2012}, \citet{fiorentino2014}, \citet{kopachi2015}, \citet{martinazzi2015,martinazzi2015b}, and \citet{lee2016}. Some derivations of the PL relations included both SXP and $\delta$~Scuti stars \citep{poretti2008,mcnamara2011}, at which two recent works suggested the PL relation for $\delta$~Scuti stars could be broken \citep{gaia2022} or segmented \citep{mv2022}. In terms of theoretical studies, \citet{santolamazza2001}, \citet{templeton2002} and \citet{fiorentino2015} presented theoretical PL relations for SXP (and $\delta$~Scuti) stars. Since SXP stars can pulsate in radial-modes (fundamental, first-overtone, and second-overtone) and/or in non-radial modes, their PL(Z) relations were primarily derived for the fundamental mode SXP stars (in some cases the first-overtone mode SXP stars were also included).

The empirical PL(Z) relations presented in the previous work were mainly in the optical $B$- and/or $V$-band, with a few exceptions such as \citet[][including $I$-band]{af2011} and \citet[][including $g$-band]{vivas2019}. Beyond optical bands, \citet{navarrte2017} derived the near-infrared $JK_s$-band PL relations based on the fundamental mode SXP stars found in $\omega$~Centauri. Nowadays, subsets of the Sloan Digital Sky Survey (SDSS) $ugriz$ filters or their variants are widely used in a variety of synoptic sky surveys. For example, \citet{vivas2020} discovered one SXP star in the Crater II dwarf galaxy using $gi$-band time-series observations taken with the Dark Energy Camera \citep[DECam,][]{flaugher2015}. In coming years, the Vera C. Rubin Observatory Legacy Survey of Space and Time \cite[LSST,][]{lsst2019} will conduct a 10-years synoptic survey in the $ugrizy$ filters. The LSST is expected to discover new dwarf galaxies and identify new SXP stars in nearby dwarf galaxies. Therefore, it is useful to derive PL relations for SXP stars in the available Sloan-like filters for their application as a distance indicators in the LSST era. 

Zwicky Transient Facility \citep[ZTF,][]{bellm2017,bellm2019,gra19} is a synoptic imaging survey on the northern sky. The ZTF observing system consists of a 600~megapixel mosaic CCD camera and the 48-inch Samuel Oschin Telescope in Schmidt design, providing a field-of-view of 47-square-degree with a pixel scale of $1.01\arcsec/$pixel \citep{dec20}. Time-series observations of ZTF were conducted in the customized $gri$ filters, and calibrated to PAN-STARRS1 \citep{chambers2016,magnier2020} AB magnitude system \citep{mas19}. Therefore, the main goal of our work is to derive the $gri$-band PL relations for SXP stars located in the northern globular clusters using ZTF light curves.

Structure of this paper is organized as follow. Section \ref{sec2} describes the sample of SXP stars and their ZTF light curves data. In Section \ref{sec3}, we refined the periods for these SXP stars and obtained their mean magnitudes based on the fitted ZTF light curves. After identifying the pulsation modes for the final selected sample of SXP stars, we derived the PL and the extinction-free period-Wesenheit (PW) relations for SXP stars in Section \ref{sec4}. We tested our PL and PW relations using one SXP star in Crater II dwarf galaxy by determining its distance in Section \ref{sec5}. The conclusions of our work are presented in Section \ref{sec6}.

\section{Sample and Data} \label{sec2}

Early compilations of SXP stars in globular clusters were presented in \citet{rl2000} and \citet{santolamazza2001}, and subsequently updated by \citet{cohen2012} with a list of 263 SXP stars. Besides the \citet{cohen2012} catalog, the ``Updated Catalog of Variable Stars in Globular Clusters'' \citep[][hereafter Clement's Catalog]{clement2001,clement2017} has also compiled a list of SXP stars in the globular clusters. After excluding those SXP stars located south of declination $-30^\circ$ (as they are outside the visibility of ZTF), there are 111 and 128 SXP stars left in the \citet{cohen2012} catalog and the Clement's Catalog, respectively, with 94 common SXP stars between these two catalogs.

We merged the SXP stars in these two catalogs, together with newly identified SXP stars in various globular clusters from recent work. These include NGC288 \citep{martinazzi2015,martinazzi2015b,lee2016}, NGC1904/M79 \citep{kopachi2015}, NGC4147 \citep{lata2019}, NGC5053 \citep{af2010}, NGC6205/M13 \citep{deras2019}, NGC6218/M12 \citep{kaluzny2015}, NGC6254/M10 \citep{salinas2016,rozyczka2018,af2020}, NGC6341/M92 \citep{yepez2020}, NGC6402/M14 \citep{yepez2022}, NGC6656/M22 \citep{rozyczka2017}, NGC6779/M56 \citep{deras2022}, NGC6934 \citep{yepez2018}, and NGC7089/M2 \citep{salinas2016}.

Furthermore, we rejected SXP stars with uncertain classification or membership, as well as those not a member of the globular clusters. The rejected SXP stars are: V83 in NGC2419 \citep{clement2017}, SX8, SX17, and SX24 in NGC5024 \citep{bramich2012}, BS19 in NGC5053 \citep{af2010}, V25 in NGC6093 \citep{clement2017}, V24 in NGC6218 \citep{sariya2018}, V8, V9, V11, and V14 in NGC6254 \citep{rozyczka2018,af2020}, V34 and V36 in NGC6341 \citep{yepez2020}, V6 in NGC6366 \citep{sariya2015}, KT-05 in NGC6656 \citep{rozyczka2017}, V11 in NGC6779 \citep{clement2017}, and QU Sge in NGC6838/M71 \citep{mccormac2014}.\footnote{QU Sge is a semi-detached binary system, where its primary is a SXP star \citep{jeon2006}, which could be a foreground star \citep{mccormac2014}.} Four SXP stars in NGC5272 (Anon, SE174, NW449, and NW858) were also removed from the sample because no accurate celestial coordinates can be found for them.

\begin{figure*}
  \gridline{\fig{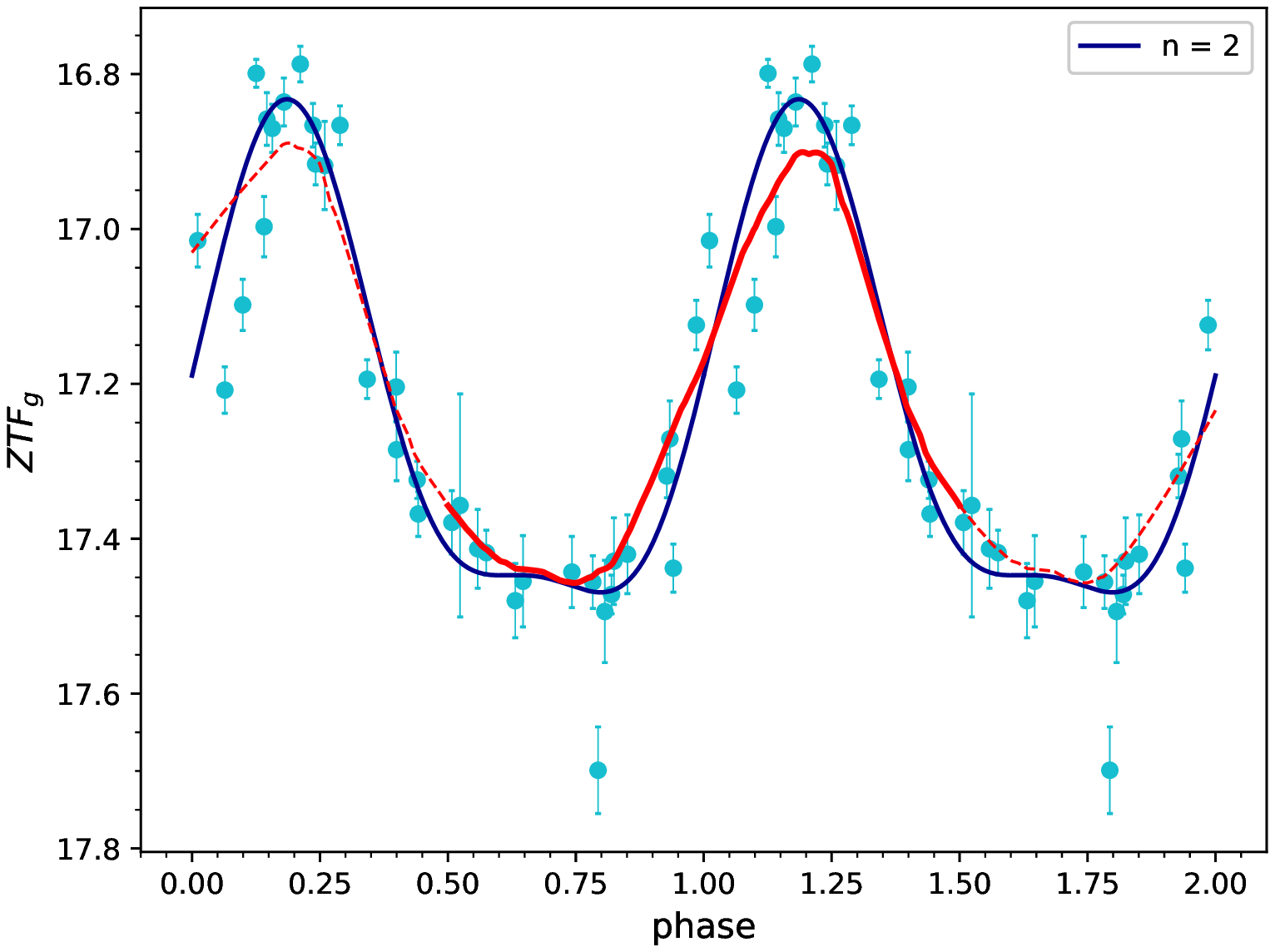}{0.32\textwidth}{}
    \fig{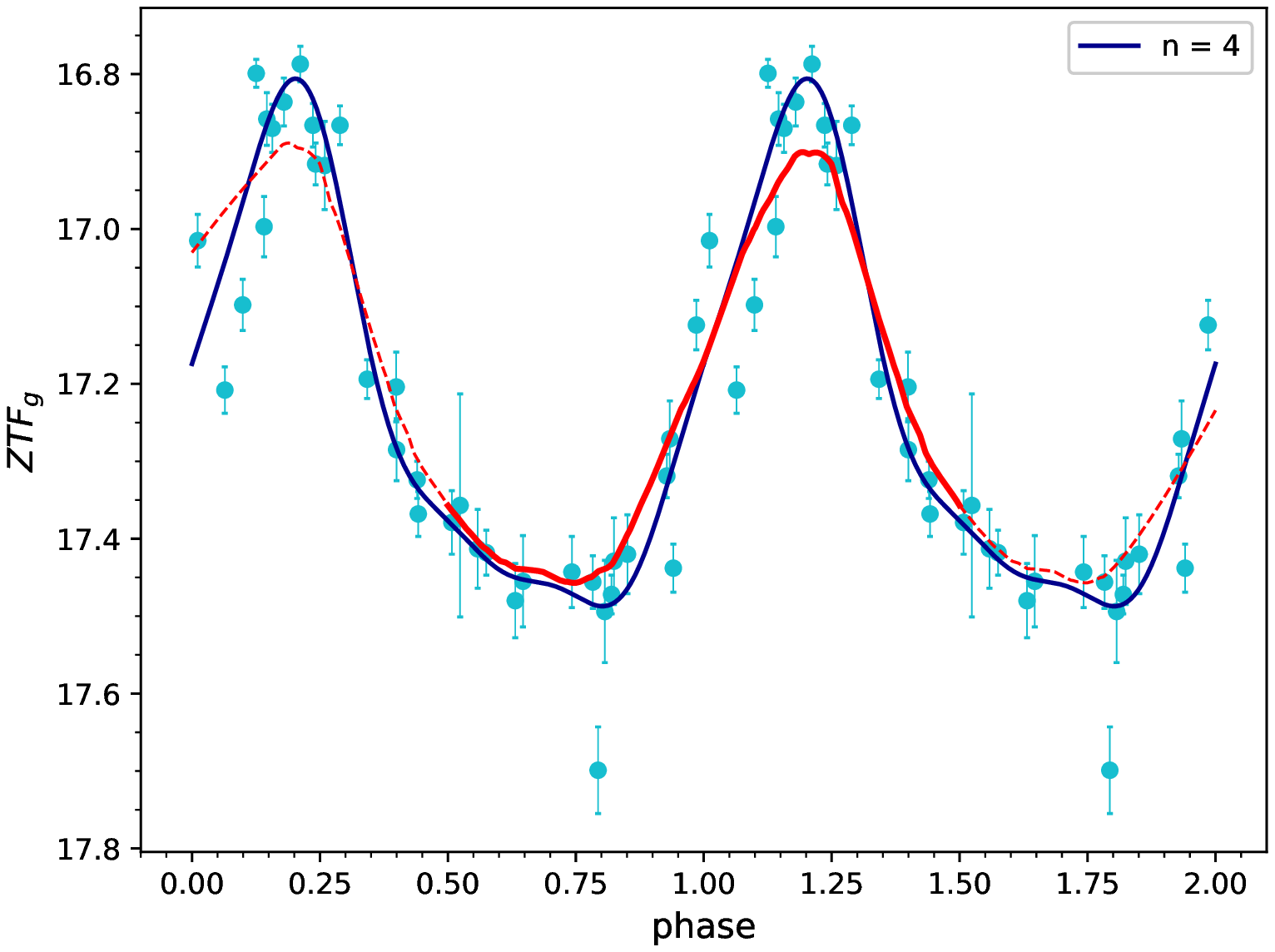}{0.32\textwidth}{}
    \fig{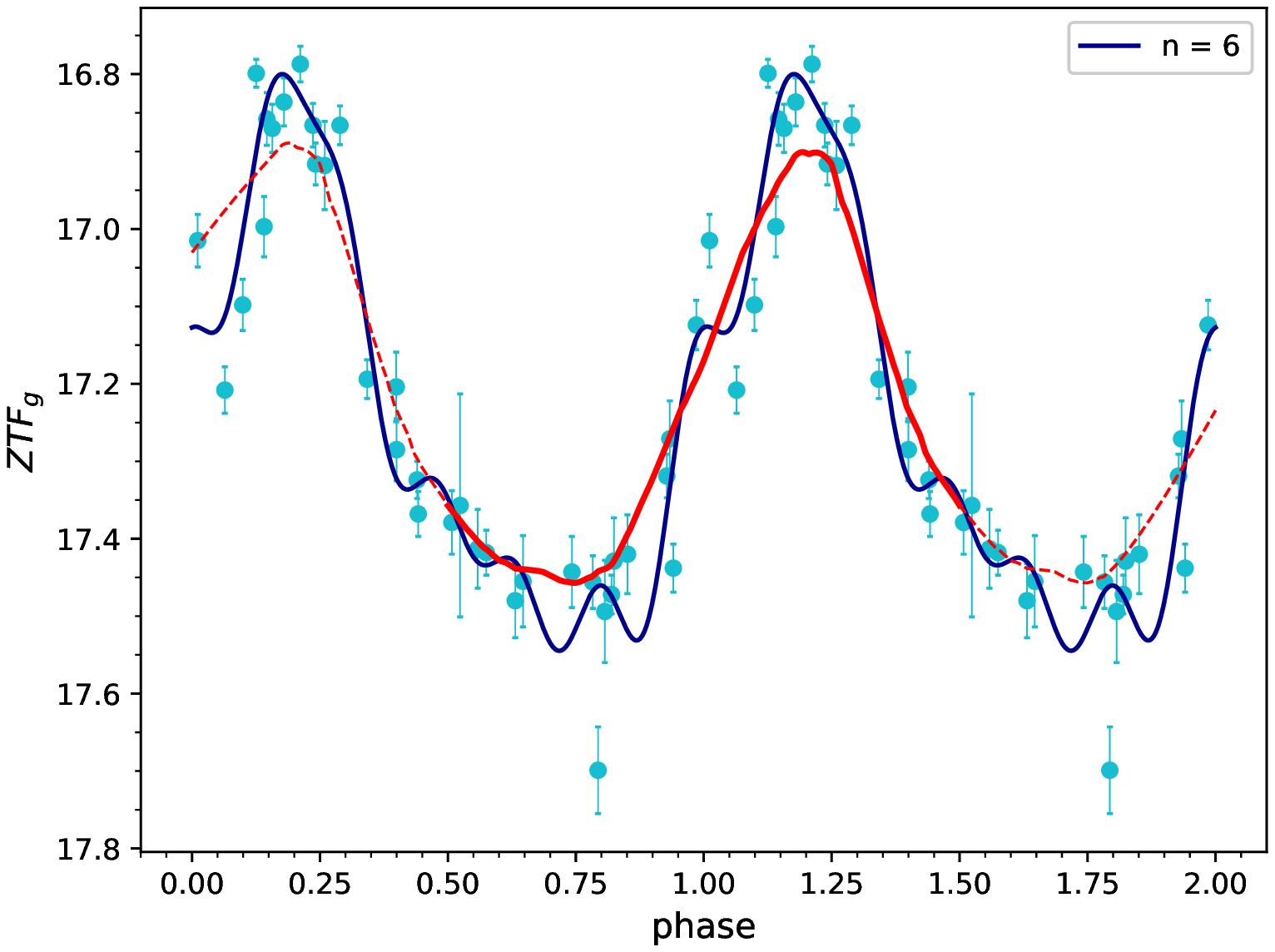}{0.32\textwidth}{}
  }
  \gridline{\fig{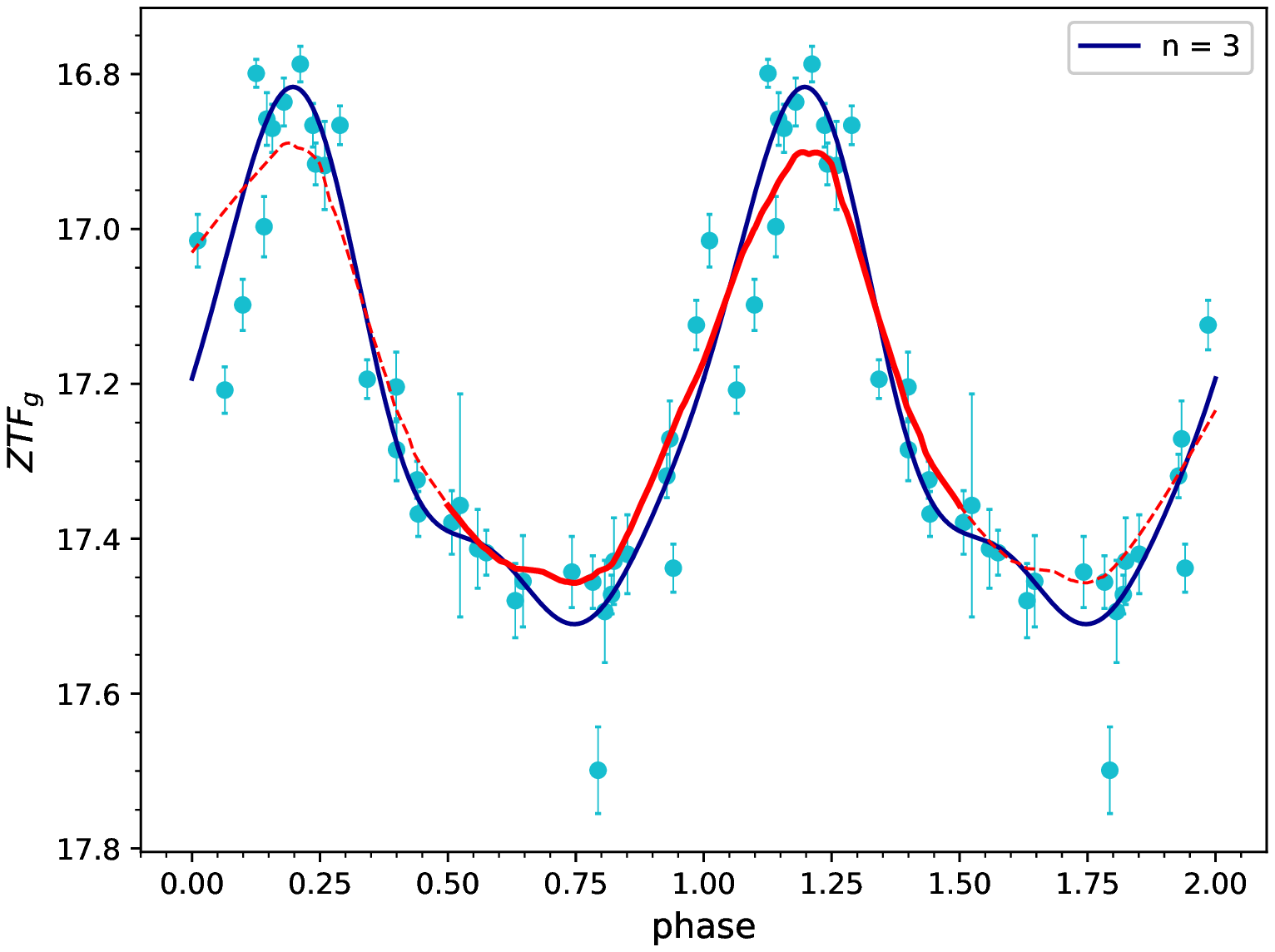}{0.32\textwidth}{}
    \fig{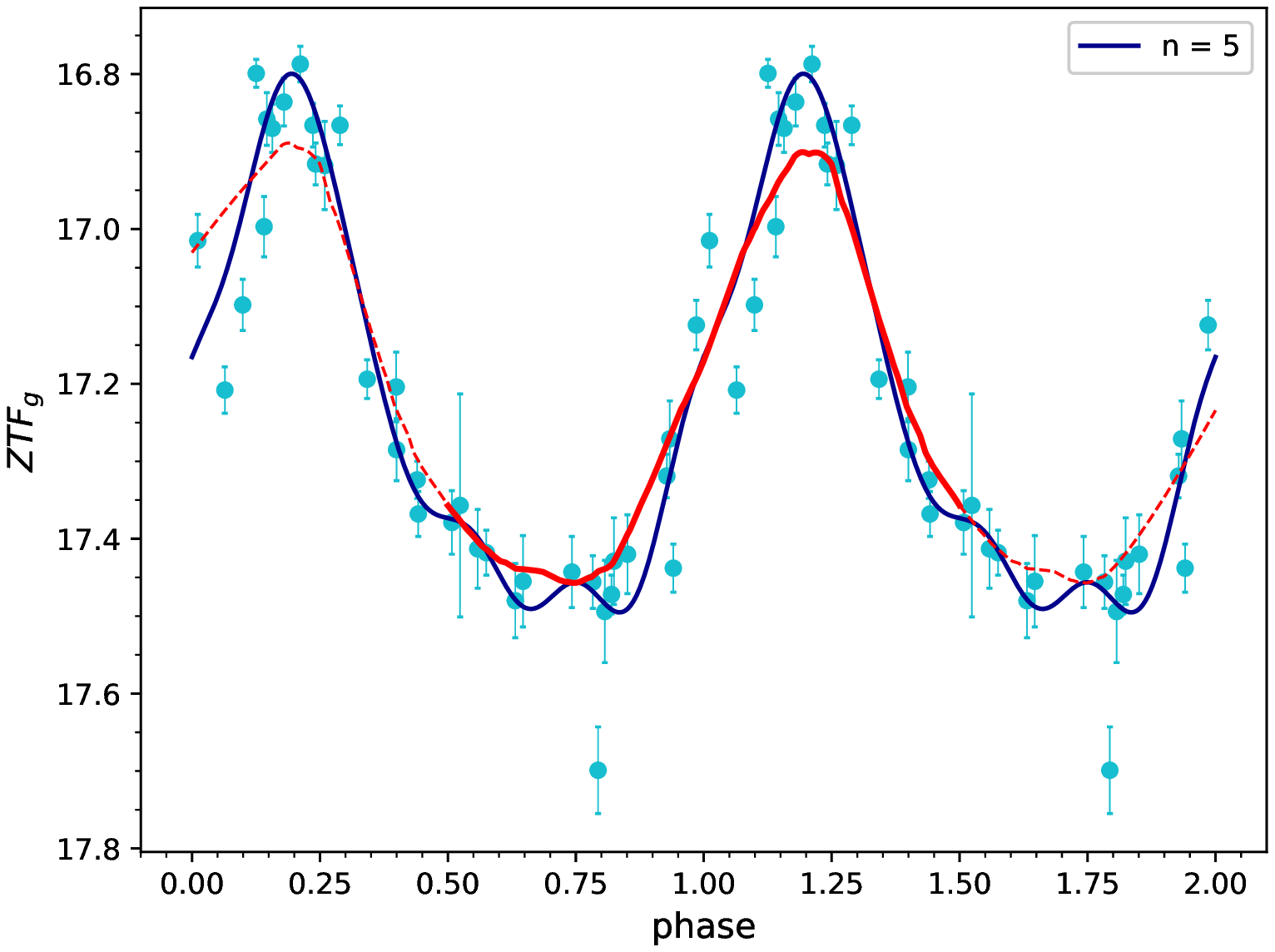}{0.32\textwidth}{}
    \fig{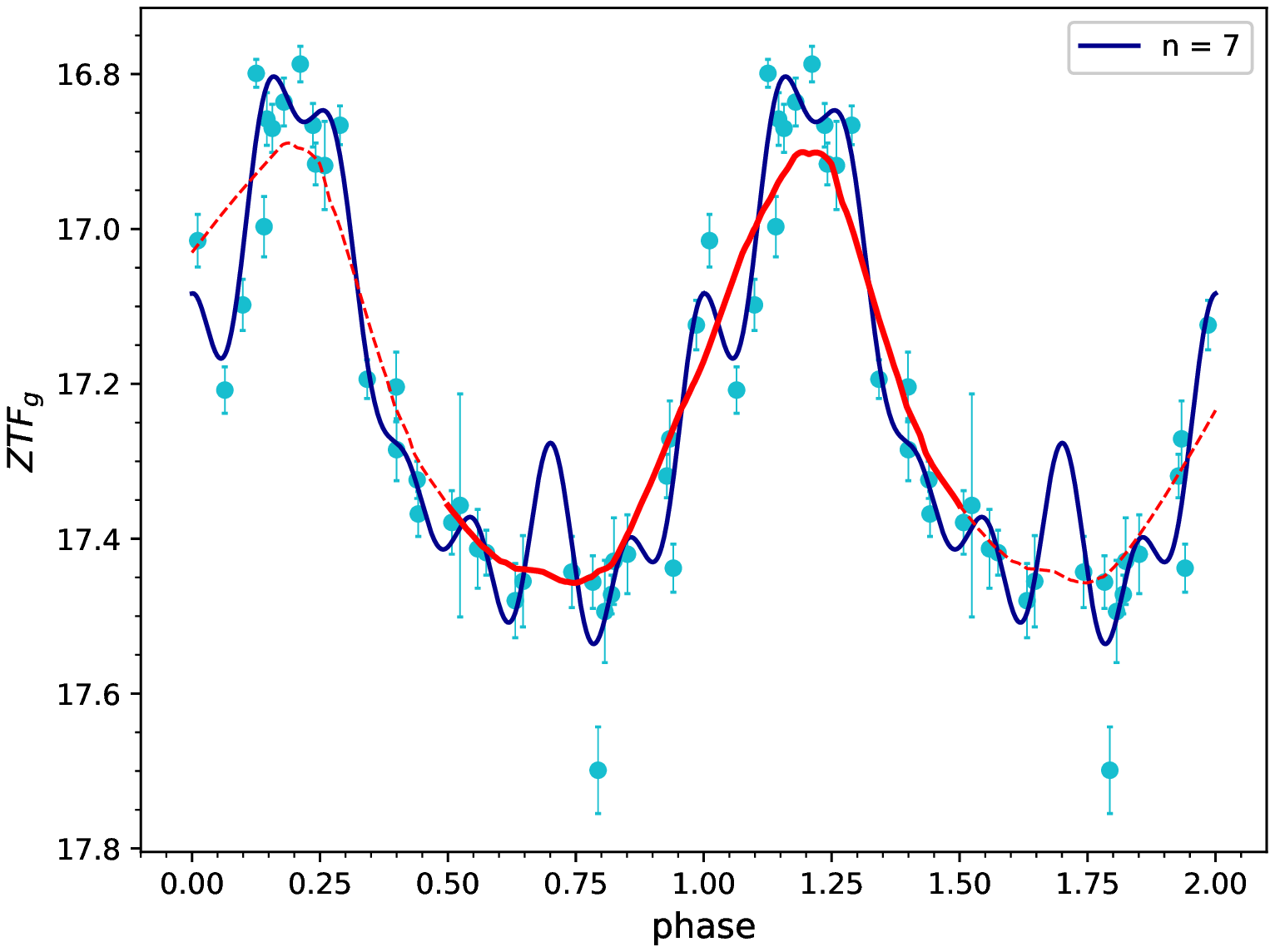}{0.32\textwidth}{}
  }
  \caption{The ZTF $g$-band observed light curve data for NGC 6254 V20 fitted with various $n^{\mathrm{th}}$-order of the Fourier expansion (dark solid curves). The Baart's condition suggested the best Fourier fit is for $n=6$. The dashed red curve is a smoothed curve derived from applying the LOWESS algorithm to the duplicated observed light curve from phase 0 to 2. The two problems mentioned in the text can be seen on the smoothed curve, such that this curve is disjointed at phase 0 and 2 (i.e. non-periodic), and a slight deviation of the smoothed curve from the data points can be seen around phase $\sim 0$. Therefore, we only compare the smoothed curve to the fitted Fourier expansion for different $n$ at phases between 0.5 and 1.5 (the solid red curve). Based on such approach, the best Fourier-order was found to be $n=4$ for this observed light curve.}
  \label{fig_examplelc}
\end{figure*}

All together, our sample contains 161 SXP stars located in 29 globular clusters. ZTF light curves for these SXP stars were extracted from the PSF (point-spread function) catalogs, produced from the dedicated ZTF reduction pipeline \citep{mas19}, using a search radius of $1\arcsec$. These PSF catalogs include those from the ZTF Public Data Release 13 and ZTF partner surveys\footnote{The high-level ZTF surveys are divided into public surveys, partner surveys and California Institute of Technology (Caltech) surveys, see \citet{bellm2019} for more details.} until 30 September 2022. Out of the 161 SXP stars in our sample, 19 of them do not have ZTF data, and 21 of them have very sparse ZTF light curve data (with total number of data points less than 20 in all $gri$ bands). These 40 SXP stars were excluded for further analysis.

\section{Periods Refinement and Light Curves Fitting} \label{sec3}

\subsection{Two-Steps Period Refining Process} \label{3.1}

Similar to our previous work \citep{ngeow2022b}, we refined the pulsation periods for the 121 SXP stars, selected in previous section with sufficient number of observations, using the {\tt LombScargleMultiband} (LSM) module \citep{vdp2015}\footnote{Available in the {\tt astroML/gatspy} package from \url{https://github.com/astroML/gatspy}.} in a two-steps process. The first step is running the LSM module on the $gri$-band (whenever available) ZTF light curves for each SXP stars within a period range of $0.005$ to $0.500$~days. We then fold the light curves based on the determined periods, and fit the folded light curves using a low order Fourier expansion \citep[for example, see][]{deb2009,bhardwaj2015,ngeow2022b}. Data points deviated more than $3s$, where $s$ is the dispersion of the fitted light curve, were discarded and the LSM module was run for the second time to determine the final adopted periods ($P_{LSM}$). 

\subsection{LOWESS-Assisted Fourier Fitting}\label{3.2}

In our previous work, we selected the best $n$-order of the Fourier expansion via visual inspection. Alternatively, the best Fourier-order can be determined using the Baart's condition \citep{petersen1986,deb2009}. However, based on our past experience, the Baart's condition might not always work. An example is shown in Figure \ref{fig_examplelc}, at which Baart's condition suggested the best Fourier order is $n=6$ for this light curve. Clearly, the fitted $6^{\mathrm{th}}$-order Fourier expansion resulted some of the small numerical bumps seen around phases of $\sim0.5$ to $\sim0.8$.

\begin{figure*}
  \gridline{\fig{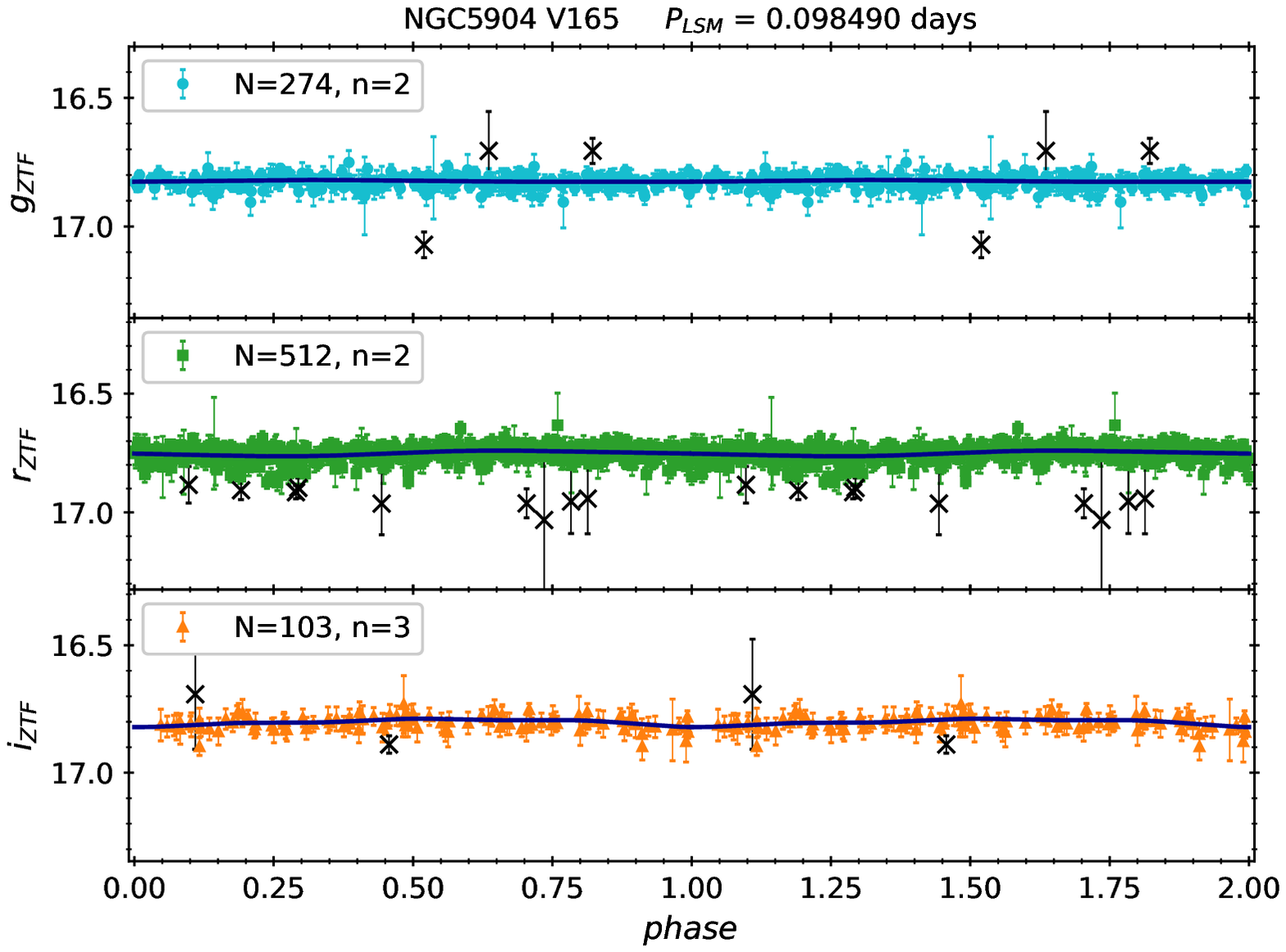}{0.32\textwidth}{}
    \fig{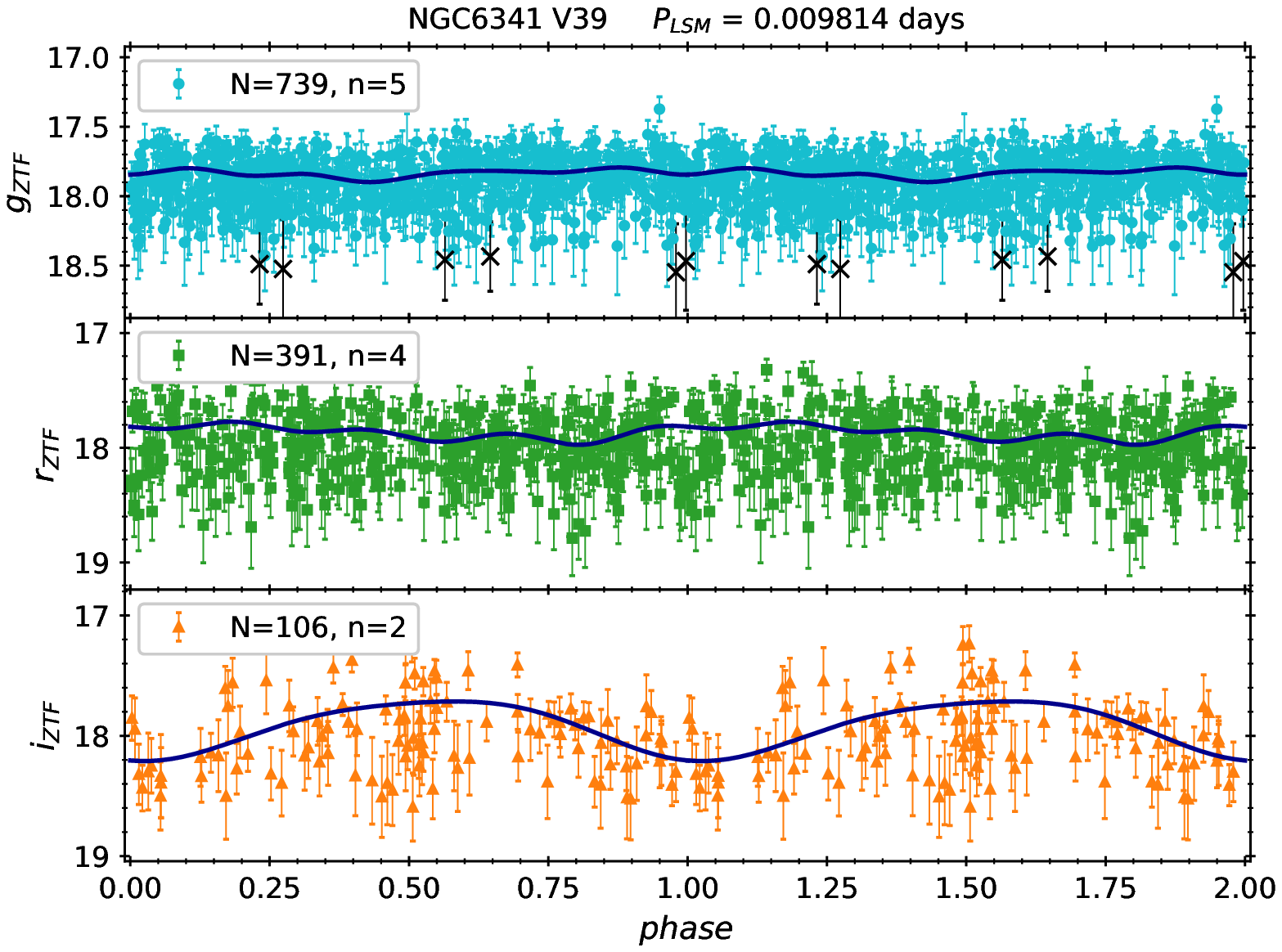}{0.32\textwidth}{}
    \fig{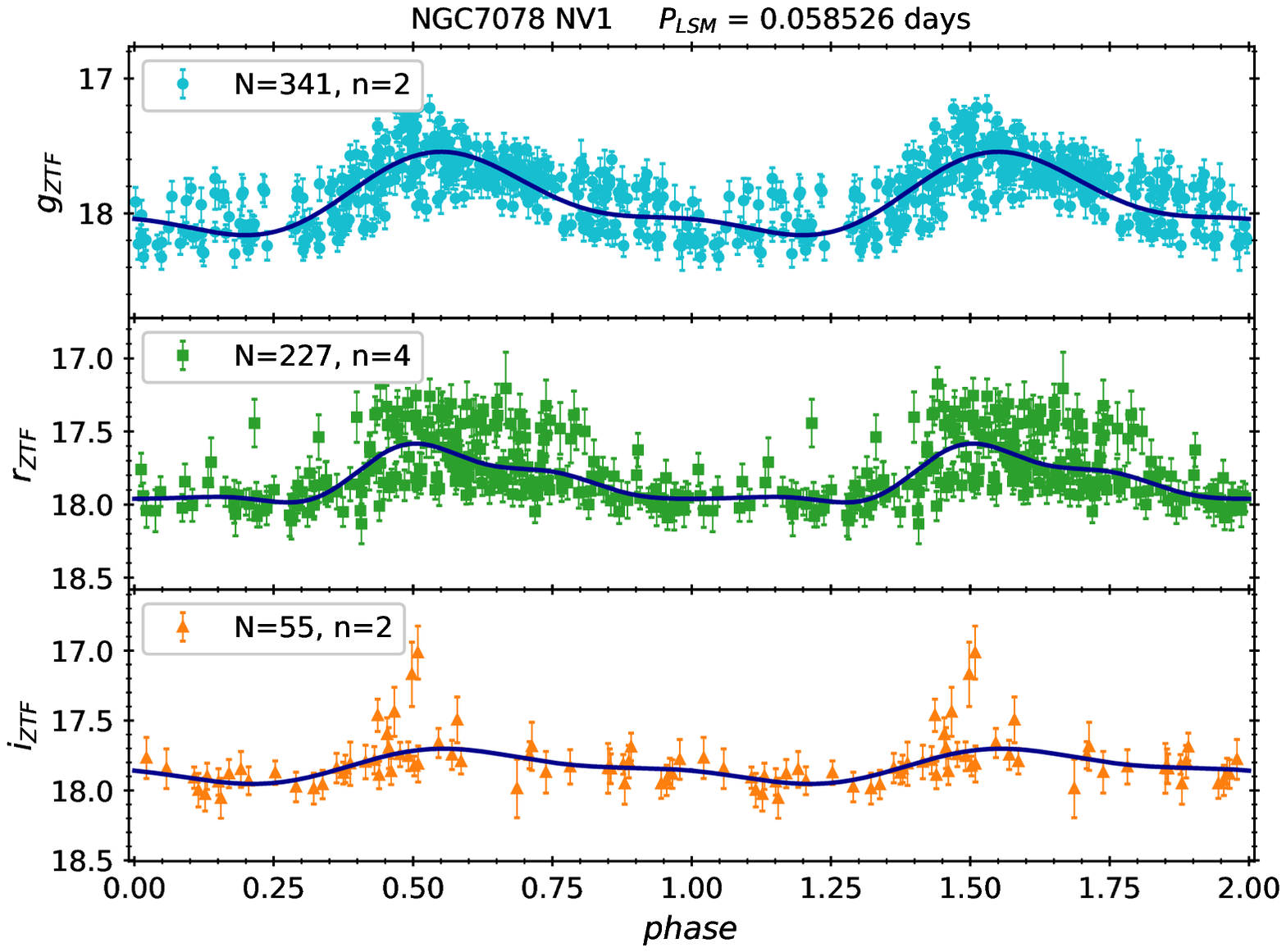}{0.32\textwidth}{}
  }
  \caption{Examples of the light curves for rejected SXP stars in our sample. $N$ represents the number of data points in the light curve. The solid curves are the fitted Fourier expansion with the best-fit Fourier order given as $n$. Black crosses are rejected data points based on the two-steps period refining process as described in Section \ref{3.1}.}
  \label{fig_bad}
\end{figure*}

\begin{figure*}
  \gridline{\fig{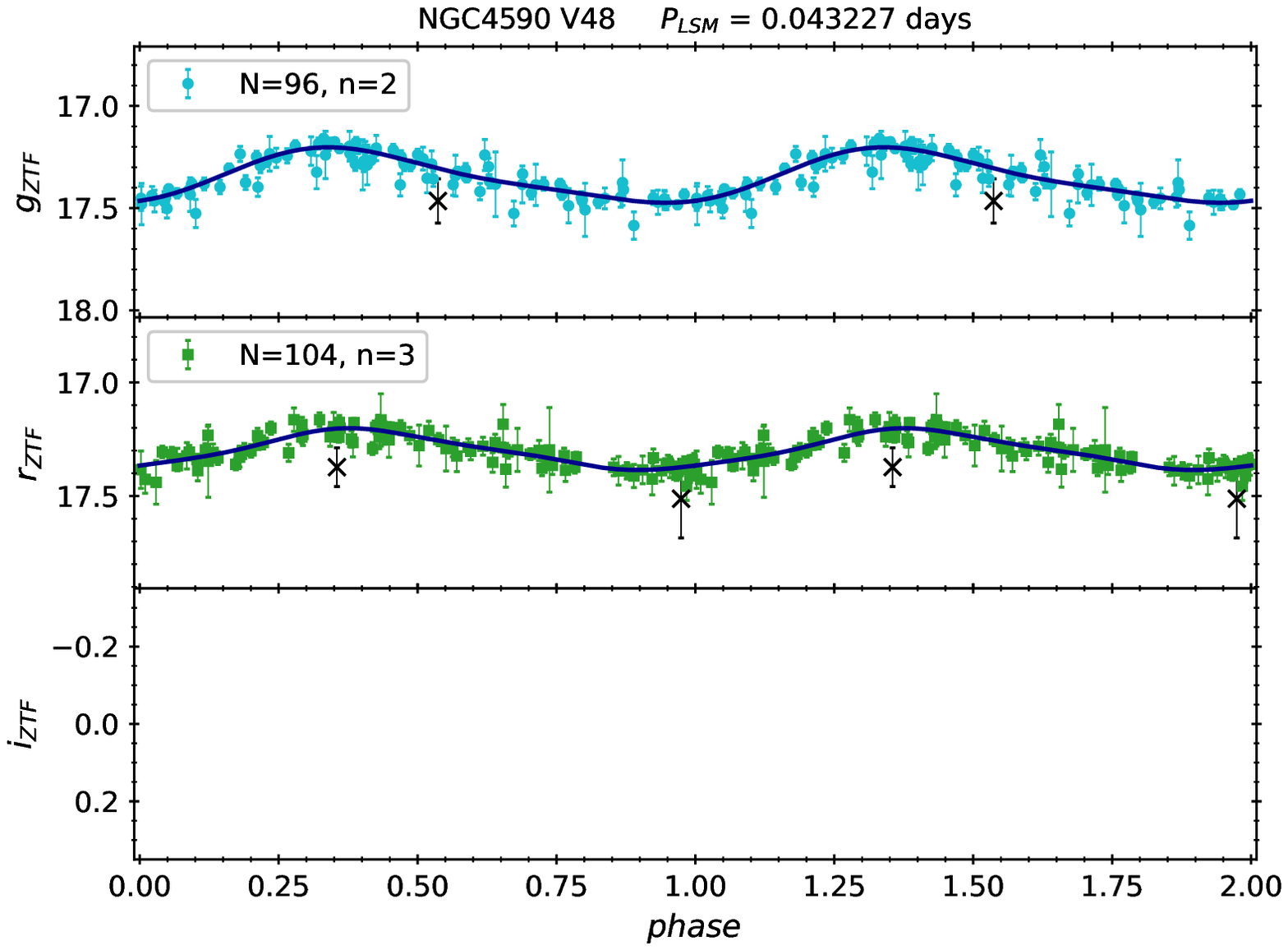}{0.32\textwidth}{}
    \fig{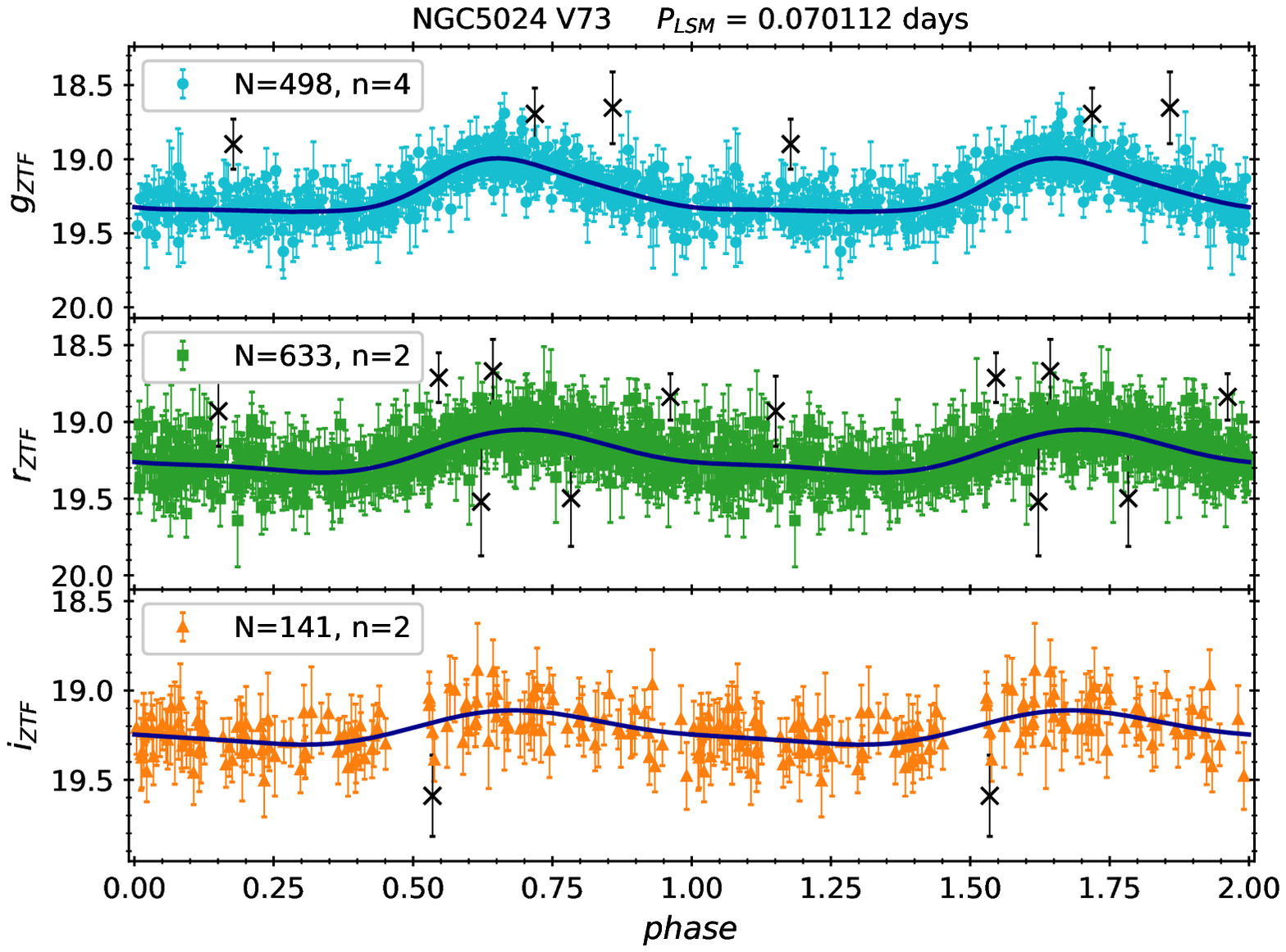}{0.32\textwidth}{}
    \fig{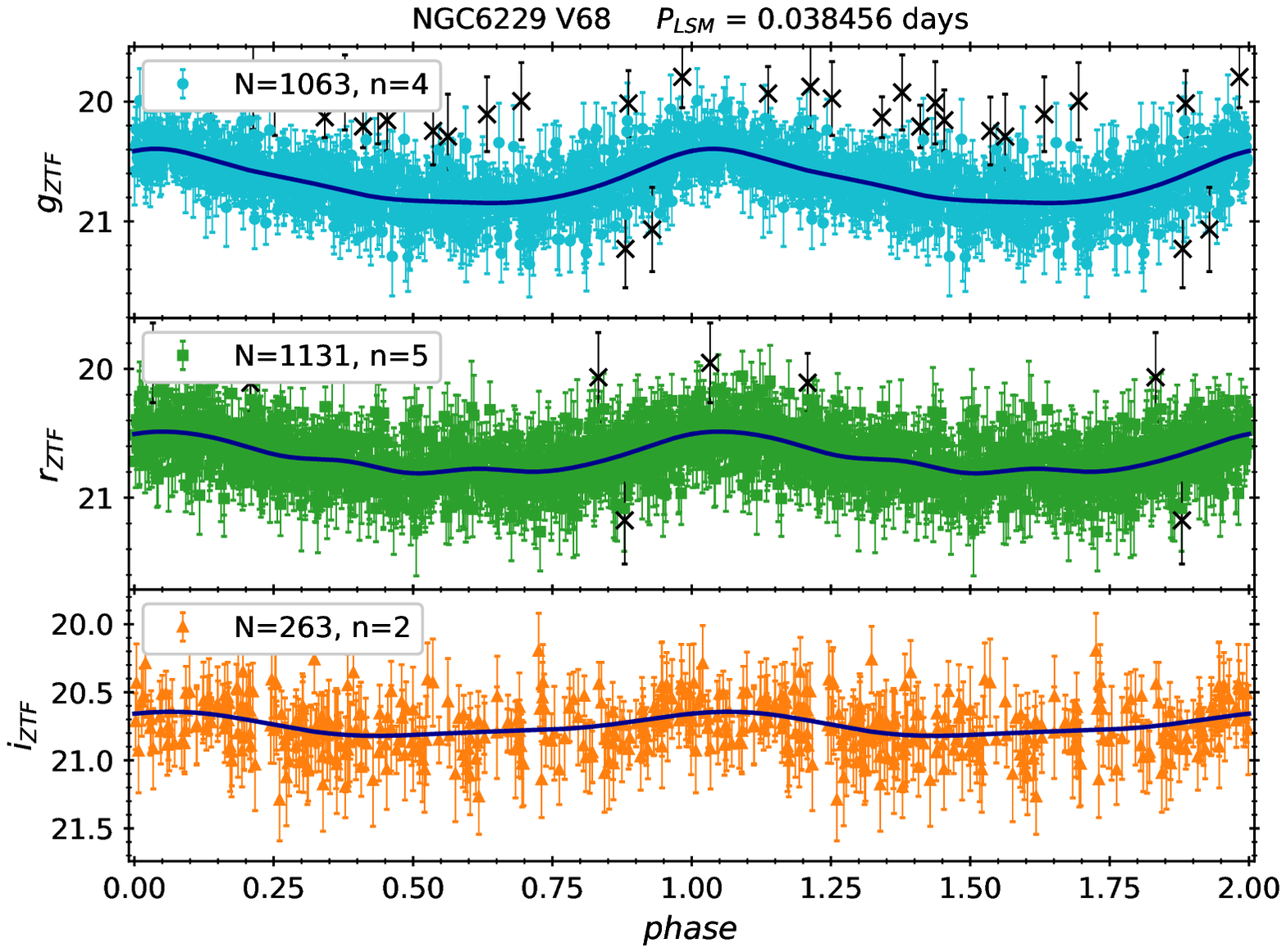}{0.32\textwidth}{}
  }
  \caption{Same as Figure \ref{fig_bad}, but for SXP stars selected in the final sample.}
  \label{fig_good}
\end{figure*}

Since the shapes of the light curves for pulsating stars are either sawtooth-like or sinusoidal, the best-fit Fourier expansion should be close to a smoothed light curve portraited by the observed data points. We employed the non-parametric LOWESS \citep[LOcally WEighted Scatterplot Smoothing,][]{cleveland1979} algorithm, implemented in {\tt statsmodels} package, to smooth the observed light curve. Nevertheless, such smoothed curve will not be continuous crossing phase 0 or 1 (i.e. non-periodic), and might not fit well the data points at the ``borders'' (i.e, at phases close to 0 or 1) of the light curve. To remedy these problems, we duplicated the observed light curves such that the observed light curve spanned from phase 0 to 2 (hence, forcing the continuity at phase 1), and applied LOWESS algorithm to this light curve. The resulted smoothed curve is shown as the dashed red curve in Figure \ref{fig_examplelc}. We then fit the original observed light curve with various low-order (typically $n=2$ to $5$) Fourier expansion. The best Fourier-order was chosen if the fitted Fourier expansion is closest to the smoothed curve within the phase of 0.5 to 1.5 (the solid red curve in Figure \ref{fig_examplelc}). For the example light curve shown in Figure \ref{fig_examplelc}, the best Fourier-order was found to be $n=4$ based on the smooth LOWESS curve. We have applied this method in our two-steps period refining process mentioned earlier. The best-fit Fourier expansion was also used to obtain the intensity mean magnitude ($\langle m \rangle$, where $m=\{g,\ r,\ i\}$) and amplitude for a given light curve.

\subsection{Selection of Final Sample}\label{3.3}

\begin{figure*}
  \gridline{\fig{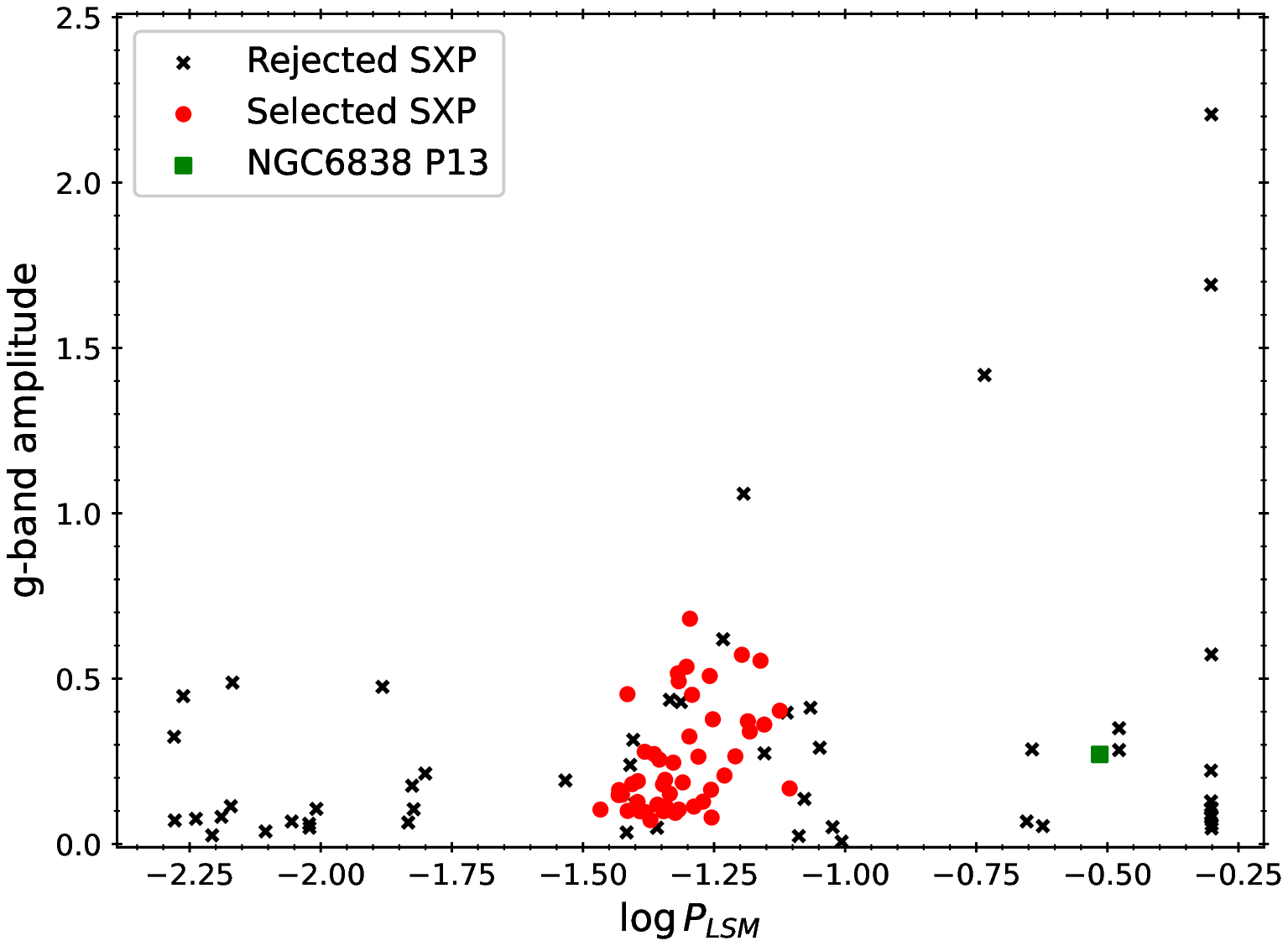}{0.32\textwidth}{}
    \fig{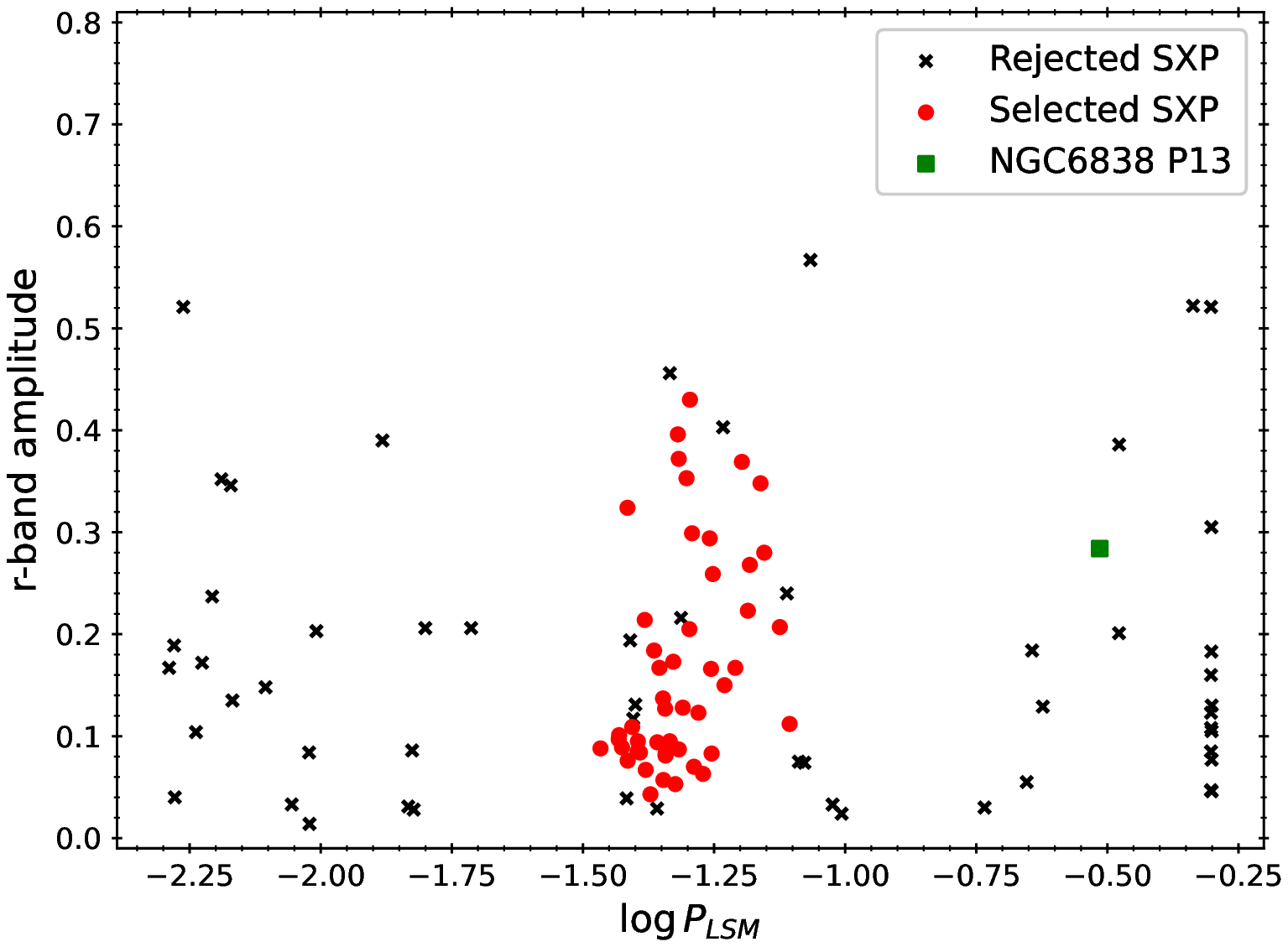}{0.32\textwidth}{}
    \fig{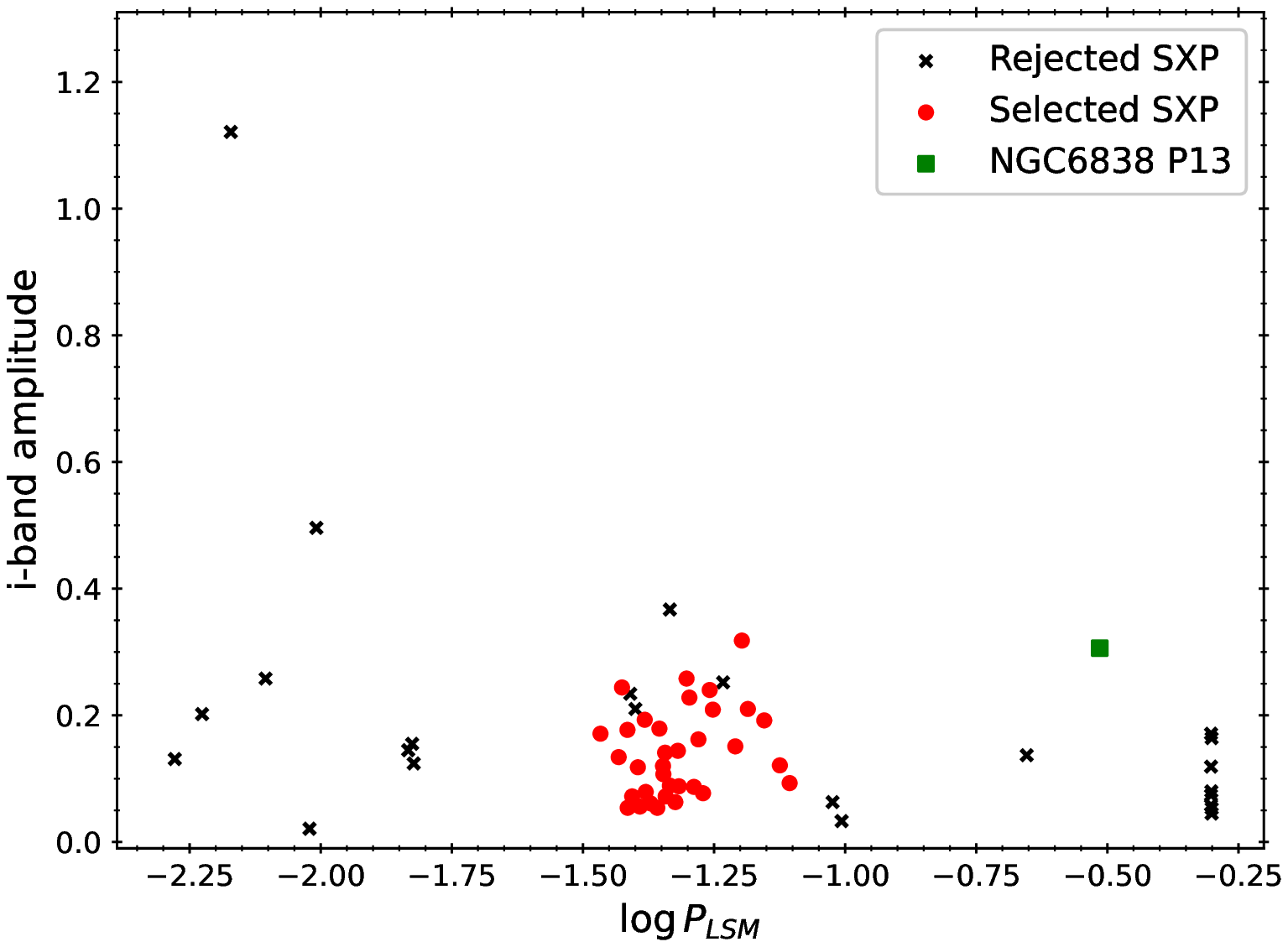}{0.32\textwidth}{}
  }
  \caption{Amplitudes as a function of $\log P_{LSM}$ for the SXP stars in the sample. See text for further discussion on NGC 6838 P13. }
  \label{fig_amp}
\end{figure*}

\begin{deluxetable*}{llllcrrrcccrr}
  %\movetableright=-1in
  \tabletypesize{\scriptsize}
  \tablecaption{Observed Properties of the Final Selected SXP Stars in Globular Clusters\label{tab1}}
  \tablewidth{0pt}
  \tablehead{
    \colhead{G. C.} &
    \colhead{Var. Name} &
    \colhead{$P_{\mathrm{lit}}$\tablenotemark{a} (days)} &
    \colhead{$P_{LSM}$ (days)} &
    \colhead{Mode\tablenotemark{b}} &
    \colhead{$N_g$} &
    \colhead{$N_r$} &
    \colhead{$N_i$} &
    \colhead{$\langle g\rangle$} &
    \colhead{$\langle r\rangle$} &
    \colhead{$\langle i\rangle$} &
    \colhead{$D$\tablenotemark{c}} &
    \colhead{$E$\tablenotemark{d}} 
  }
  \startdata
NGC0288 & V5	& 0.05107 & 0.05106683 & 1O	&184	& 180	& 11	& 17.475 & 17.524 & $\cdots$ & $8.99\pm0.09$ & $0.016\pm0.002$ \\
NGC4590 & V48	& 0.04320 & 0.04322702 & 2O	&96	& 104	& 0	& 17.344 & 17.298 & $\cdots$ & $10.40\pm0.10$ & $0.000\pm0.000$ \\
NGC4590 & V50	& 0.06580 & 0.06581527 & 1O	&89	& 94	& 0	& 17.572 & 17.436 & $\cdots$ & $10.40\pm0.10$ & $0.000\pm0.000$ \\
NGC5024 & V73	& 0.07010 & 0.07011209 & F	&498	& 633	& 141	& 19.228 & 19.212 & 19.218 & $18.50\pm0.18$ & $0.000\pm0.000$ \\
NGC5024 & V74	& 0.04540 & 0.04537313 & 1O	&874	& 1218	& 245	& 19.146 & 19.161 & 19.213 & $18.50\pm0.18$ & $0.000\pm0.000$ \\
NGC5024 & V75	& 0.04430 & 0.04424759 & F	&823	& 1100	& 225	& 19.549 & 19.518 & 19.556 & $18.50\pm0.18$ & $0.000\pm0.000$ \\
NGC5024 & V76	& 0.04150 & 0.04148953 & F	&718	& 898	& 187	& 19.777 & 19.840 & 19.914 & $18.50\pm0.18$ & $0.000\pm0.000$ \\
NGC5024 & V77	& 0.07690 & 0.07838129 & 1O	&931	& 1330	& 262	& 18.558 & 18.496 & 18.544 & $18.50\pm0.18$ & $0.000\pm0.000$ \\
NGC5024 & V79	& 0.04630 & 0.04631969 & 2O	&910	& 1298	& 257	& 18.832 & 18.687 & 18.713 & $18.50\pm0.18$ & $0.036\pm0.002$ \\
NGC5024 & V89	& 0.04340 & 0.04547359 & F	&807	& 1091	& 222	& 19.597 & 19.597 & 19.653 & $18.50\pm0.18$ & $0.000\pm0.000$ \\
NGC5024 & V90	& 0.03850 & 0.03849888 & 1O	&900	& 1231	& 250	& 19.107 & 19.112 & 19.180 & $18.50\pm0.18$ & $0.000\pm0.000$ \\
NGC5024 & V93	& 0.04010 & 0.04254658 & 2O	&910	& 1277	& 256	& 18.777 & 18.733 & 18.802 & $18.50\pm0.18$ & $0.000\pm0.000$ \\
NGC5024 & V96	& 0.03930 & 0.03926814 & F	&791	& 1049	& 220	& 19.540 & 19.479 & 19.545 & $18.50\pm0.18$ & $0.002\pm0.002$ \\
NGC5024 & V100	& 0.04820 & 0.04819481 & F	&886	& 1197	& 243	& 19.319 & 19.256 & 19.332 & $18.50\pm0.18$ & $0.002\pm0.002$ \\
NGC5024 & V101	& 0.05250 & 0.05251568 & F	&573	& 759	& 154	& 19.289 & 19.314 & 19.310 & $18.50\pm0.18$ & $0.000\pm0.000$ \\
NGC5053 & BS28	& 0.04547 & 0.04502879 & F	&415	& 546	& 112	& 19.313 & 19.303 & 19.392 & $17.54\pm0.23$ & $0.000\pm0.000$ \\
NGC5053 & V11	& 0.03700 & 0.03700038 & 1O	&444	& 580	& 118	& 19.379 & 19.347 & 19.430 & $17.54\pm0.23$ & $0.000\pm0.000$ \\
NGC5053 & V12	& 0.03765 & 0.03754719 & F	&441	& 565	& 107	& 19.598 & 19.610 & 19.689 & $17.54\pm0.23$ & $0.000\pm0.000$ \\
NGC5053 & V13	& 0.03396 & 0.03416461 & F	&385	& 456	& 93	& 19.630 & 19.668 & 19.791 & $17.54\pm0.23$ & $0.000\pm0.000$ \\
NGC5466 & V31	& 0.04030 & 0.04026386 & 1O	&424	& 655	& 179	& 18.802 & 18.763 & 18.843 & $16.12\pm0.16$ & $0.000\pm0.000$ \\
NGC5466 & V32	& 0.04500 & 0.04496170 & F	&363	& 563	& 152	& 19.252 & 19.278 & 19.401 & $16.12\pm0.16$ & $0.000\pm0.000$ \\
NGC5466 & V33	& 0.04990 & 0.04985460 & 1O	&415	& 618	& 159	& 18.846 & 18.803 & 18.916 & $16.12\pm0.16$ & $0.000\pm0.000$ \\
NGC5466 & V34	& 0.05070 & 0.05149370 & 1O	&444	& 686	& 185	& 18.714 & 18.679 & 18.755 & $16.12\pm0.16$ & $0.000\pm0.000$ \\
NGC5466 & V35	& 0.05050 & 0.05047364 & F	&428	& 653	& 174	& 19.078 & 19.047 & 19.138 & $16.12\pm0.16$ & $0.000\pm0.000$ \\
NGC5466 & V36	& 0.05520 & 0.05518525 & 1O	&366	& 596	& 174	& 18.724 & 18.651 & 18.691 & $16.12\pm0.16$ & $0.000\pm0.000$ \\
NGC5466 & V39	& 0.04800 & 0.04798352 & F	&343	& 338	& 36	& 19.087 & 19.079 & 19.138 & $16.12\pm0.16$ & $0.000\pm0.000$ \\
NGC5897 & V9	& 0.05060 & 0.04816860 & F	&88	& 99	& 0	& 18.814 & 18.721 & $\cdots$ & $12.55\pm0.24$ & $0.096\pm0.002$ \\
NGC5904 & V166	& 0.04170 & 0.04167214 & 1O	&381	& 744	& 132	& 17.099 & 17.112 & 17.195 & $7.48\pm0.06$ & $0.004\pm0.002$ \\
NGC5904 & V167	& 0.04740 & 0.04745436 & 2O	&400	& 769	& 135	& 17.009 & 16.970 & 17.026 & $7.48\pm0.06$ & $0.074\pm0.002$ \\
NGC6171 & V26	& 0.05260 & 0.05551172 & F	&116	& 169	& 0	& 18.090 & 17.625 & $\cdots$ & $5.63\pm0.08$ & $0.430\pm0.000$ \\
NGC6205 & V47	& 0.06526 & 0.06525557 & F	&1255	& 1137	& 192	& 17.224 & 17.129 & 17.193 & $7.42\pm0.08$ & $0.000\pm0.000$ \\
NGC6205 & V50	& 0.06175 & 0.06175429 & F	&1290	& 1308	& 254	& 17.052 & 16.969 & 16.993 & $7.42\pm0.08$ & $0.000\pm0.000$ \\
NGC6218 & V25	& 0.04903 & 0.04903408 & F	&191	& 335	& 4	& 17.176 & 16.988 & $\cdots$ & $5.11\pm0.05$ & $0.194\pm0.006$ \\
NGC6229 & V68	& 0.03850 & 0.03845637 & F	&1063	& 1131	& 263	& 20.657 & 20.675 & 20.744 & $30.11\pm0.47$ & $0.004\pm0.002$ \\
NGC6254 & V17	& 0.03695 & 0.05564910 & F	&52	& 108	& 1	& 17.601 & 17.331 & $\cdots$ & $5.07\pm0.06$ & $0.272\pm0.002$ \\
NGC6254 & V20	& 0.05060 & 0.05060309 & F	&35	& 82	& 1	& 17.213 & 17.008 & $\cdots$ & $5.07\pm0.06$ & $0.226\pm0.007$ \\
NGC6341 & V33	& 0.07509 & 0.07509381 & 2O	&1486	& 1407	& 420	& 16.455 & 16.169 & 16.045 & $8.50\pm0.07$ & $0.000\pm0.000$ \\
NGC6341 & V41	& 0.05595 & 0.05594717 & 1O	&1633	& 1597	& 454	& 17.147 & 17.132 & 17.214 & $8.50\pm0.07$ & $0.000\pm0.000$ \\
NGC6402 & V177	& 0.06898 & 0.06898326 & F	&188	& 549	& 1	& 19.469 & 18.782 & $\cdots$ & $9.14\pm0.25$ & $0.560\pm0.003$ \\
NGC6656 & KT-04	& 0.03600 & 0.03707136 & 1O	&100	& 601	& 0	& 16.948 & 16.629 & $\cdots$ & $3.30\pm0.04$ & $0.374\pm0.002$ \\
NGC6656 & KT-28	& 0.05500 & 0.05888543 & 1O	&88	& 552	& 0	& 16.447 & 16.033 & $\cdots$ & $3.30\pm0.04$ & $0.338\pm0.002$ \\
NGC6779 & V15	& 0.04552 & 0.04545172 & 1O	&363	& 656	& 24	& 18.752 & 18.421 & $\cdots$ & $10.43\pm0.14$ & $0.180\pm0.000$ \\
NGC6934 & V52	& 0.06356 & 0.06356308 & F	&258	& 562	& 117	& 19.033 & 18.870 & 18.889 & $15.72\pm0.17$ & $0.080\pm0.003$ \\
NGC6934 & V92	& 0.04586 & 0.04384961 & F	&199	& 463	& 92	& 19.607 & 19.442 & 19.494 & $15.72\pm0.17$ & $0.102\pm0.002$ \\
NGC6981 & V55	& 0.04700 & 0.04703276 & F	&196	& 228	& 0	& 19.649 & 19.609 & $\cdots$ & $16.66\pm0.18$ & $0.038\pm0.002$ \\
NGC7078 & V156	& 0.04060 & 0.04062126 & F	&701	& 805	& 181	& 18.491 & 18.487 & 18.539 & $10.71\pm0.10$ & $0.052\pm0.005$ \\
NGC7099 & V20	& 0.04020 & 0.04019862 & F	&165	& 168	& 0	& 17.784 & 17.747 & $\cdots$ & $8.46\pm0.09$ & $0.000\pm0.000$ \\
NGC7492 & V6	& 0.05660 & 0.05359233 & 2O	&176	& 203	& 43	& 19.157 & 19.034 & 19.068 & $24.39\pm0.57$ & $0.038\pm0.002$ \\
  \enddata
  \tablenotetext{a}{Published period as given in the literature.}
  \tablenotetext{b}{Pulsation mode based on the $V$-band PL relation, see Section \ref{sec4.1} for more details.}
  \tablenotetext{c}{Distance of the host globular clusters in kpc, adopted from \citet{baumgardt2021}.}
  \tablenotetext{d}{Reddening value returned from the {\tt Bayerstar2019} 3D reddening map \citep{green2019} at the location of the SXP stars in the globular clusters. Extinction corrections on each filters are: $A_g = 3.518E$, $A_r = 2.617E$ and $A_i=1.971E$ \citep{green2019}.}
\end{deluxetable*}

\begin{figure*}
  \plottwo{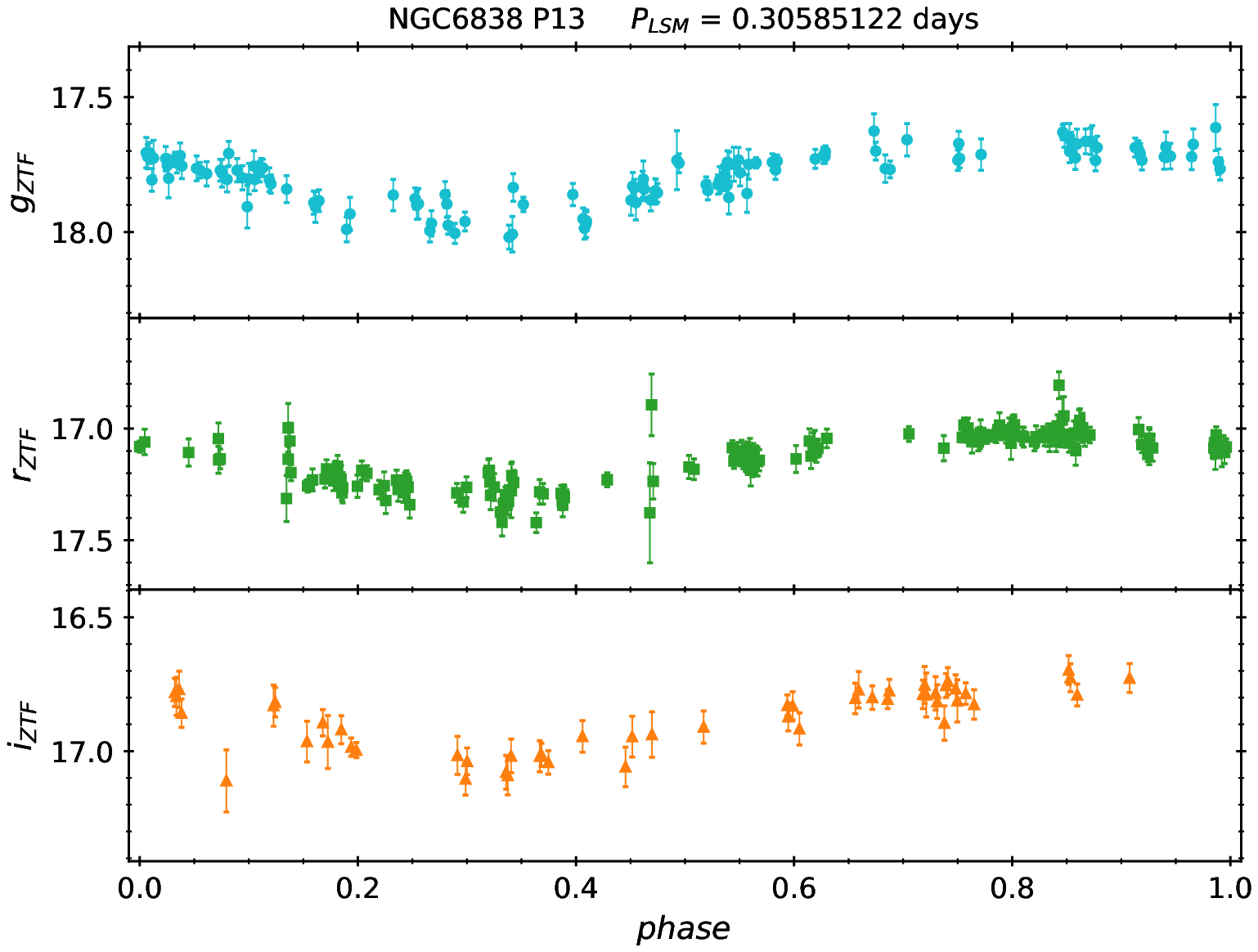}{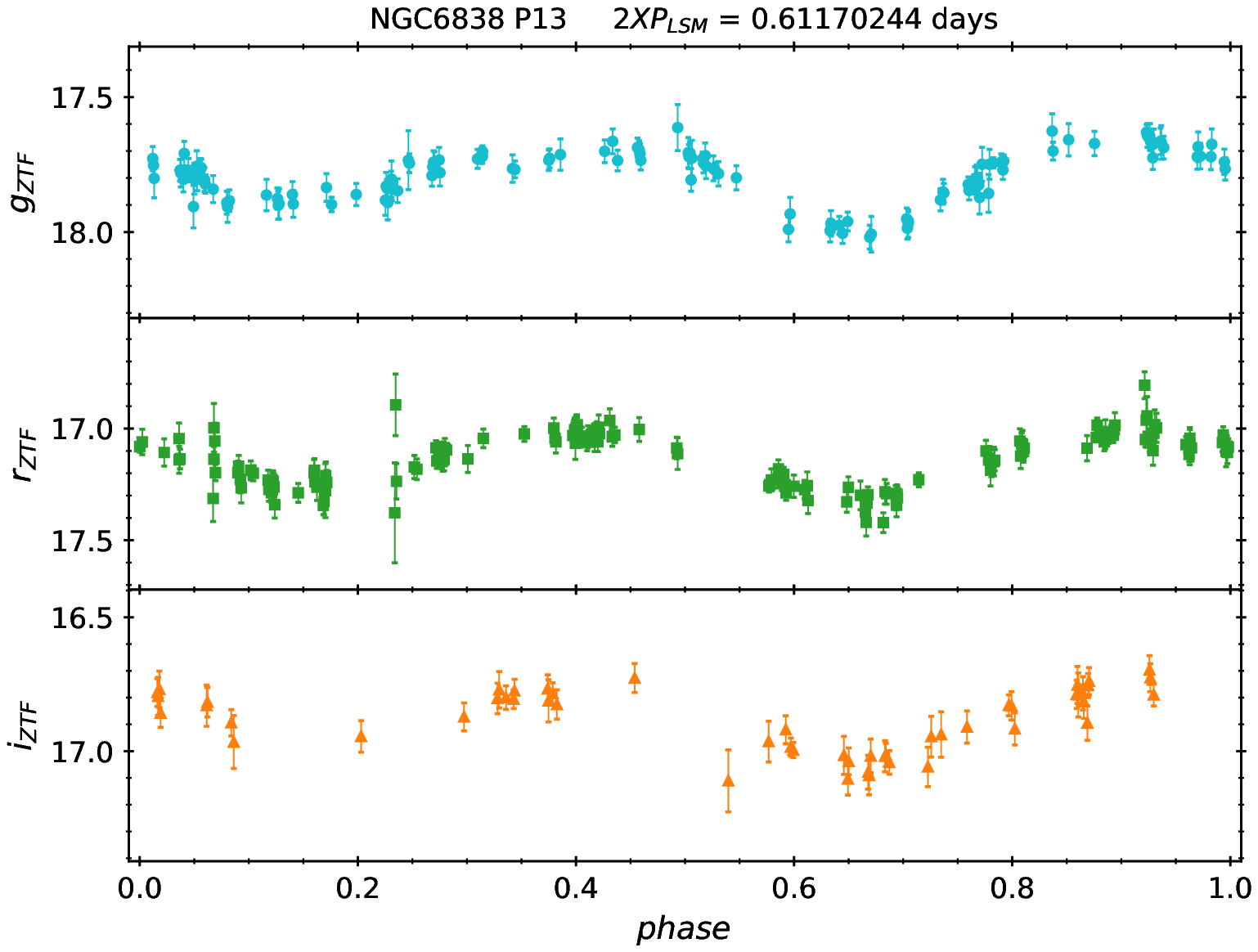}
  \caption{ZTF light curves for NGC 6838 P 13 folded with $P_{LSM}$ (left panel) and twice of $P_{LSM}$ (right panel).}
  \label{fig_p13}
\end{figure*}

Given the crowded nature of globular clusters, it is unavoidable that some of the ZTF light curves will be affected by blending. We visually inspected all of the ZTF light curves and looked for light curves that showing signs of blending, such as the light curve is flat, or the light curves exhibit too much scatter. After visual inspections, we rejected $\sim 60\%$ of the SXP stars in our samples. Some of these light curves were also rejected due to reasons other than blending, including the periods found were wrong, the amplitudes unusually large, etc. ZTF light curves for few examples of the rejected SXP stars are displayed in Figure \ref{fig_bad}. Therefore, we finally selected 49 SXP stars in our sample, some examples of their ZTF light curves are shown in Figure \ref{fig_good}. Table \ref{tab1} summarized the observed properties of these selected SXP stars.

Figure \ref{fig_amp} presents the fitted amplitudes as a function of logarithmic $P_{LSM}$. As can be seen from this Figure, the visually selected SXP stars occupy a rather narrow range in $\log P_{LSM}$ from $\sim -1.0$ to $\sim -1.5$. On the other hand, the rejected SXP stars spanned a wide range in $\log P_{LSM}$, including some of them near the boundaries of $0.005$~days ($\log P_{LSM}=-2.3$) and $0.500$~days ($\log P_{LSM}=-0.3$) when performing the period refinement using the LSM module. These periods were incorrect suggesting the ZTF light curves on these SXP stars could be affected by blending, hence they should be excluding from the sample. The only exception is NGC 6838 P13 (marked as a green filled square in Figure \ref{fig_amp}), at which its ZTF light curves are fine but it has a much longer period than other selected SXP stars.

ZTF light curves for NGC 6838 P13, folded with $P_{LSM}=0.30585$~days found in Section \ref{3.1}, are shown in the left panel of Figure \ref{fig_p13}. This star has the longest period in the \citet{cohen2012} catalog, however in its discovery paper \citep{park2000} it was commented as a possible SXP star. Indeed, this star was classified as an eclipsing binary (ECL) in Gaia Data Release (DR) 3 variability catalog \citep[][with Gaia DR3 ID of 1821624895912072576]{eyer2022}. The Gaia DR3 eclipsing binary catalog \citep{mowlavi2022} gives an orbital period of 0.61171621~days for this star, which is close to $2\times P_{LSM}$. Alternate minima can be seen from the ZTF light curves folded with $2\times P_{LSM}$, confirming the eclipsing binary nature of this star (see right panel of Figure \ref{fig_p13}) and hence it was removed from the sample. 

\begin{figure}
  \epsscale{1.05}
  \plotone{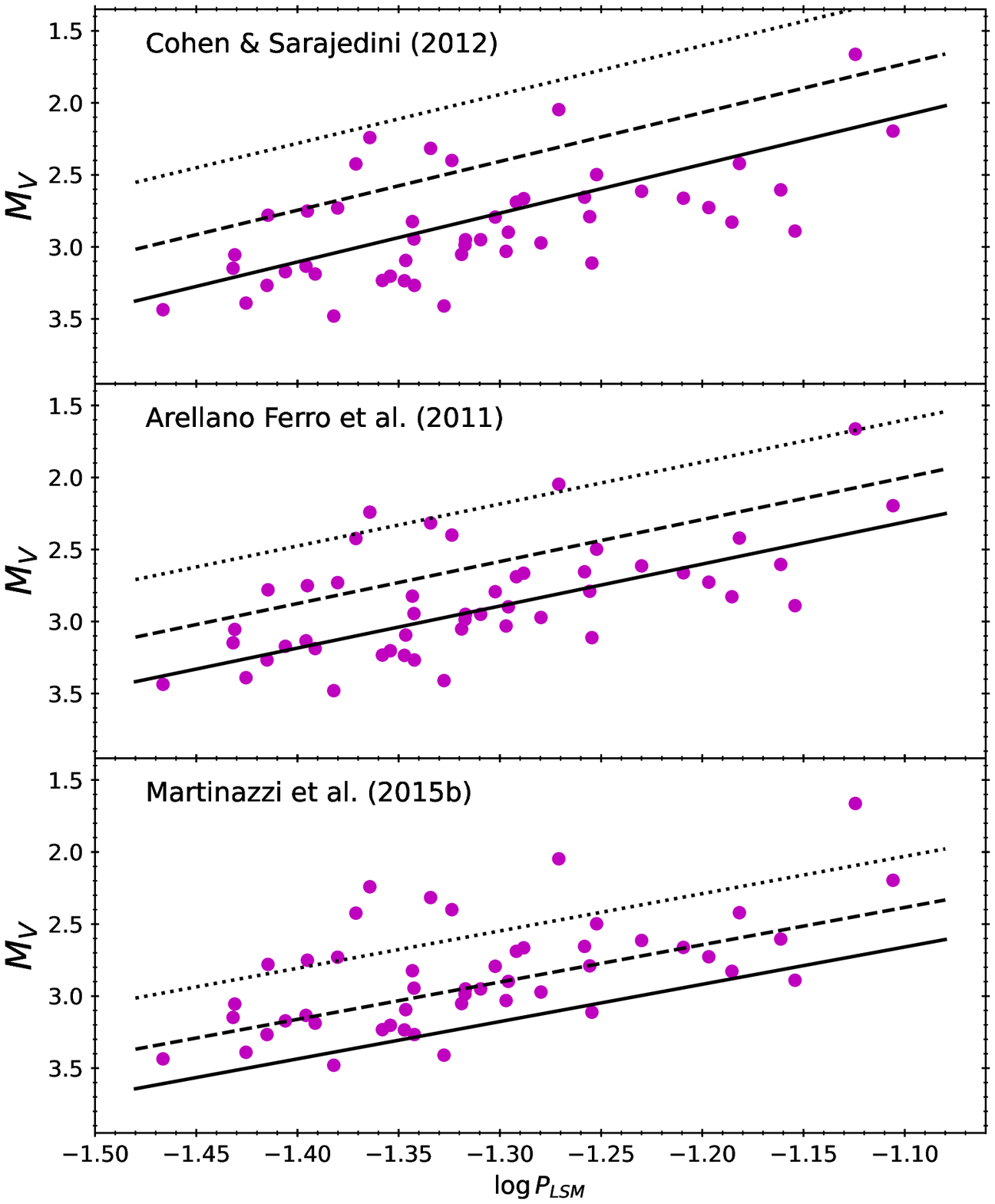}
  \caption{Comparison of the observed $V$-band PL relations for the 48 SXP stars listed in Table \ref{tab1} to the published PL relations for fundamental mode pulsators (solid lines) in the literature. The dashed and dotted lines are the corresponding PL relations for the first-overtone and the second-overtone pulsators, respectively (see text for more details).}
  \label{fig_plv}
\end{figure}

\section{The PL and PW Relations} \label{sec4}

\begin{figure*}
  \epsscale{1.05}
  \plottwo{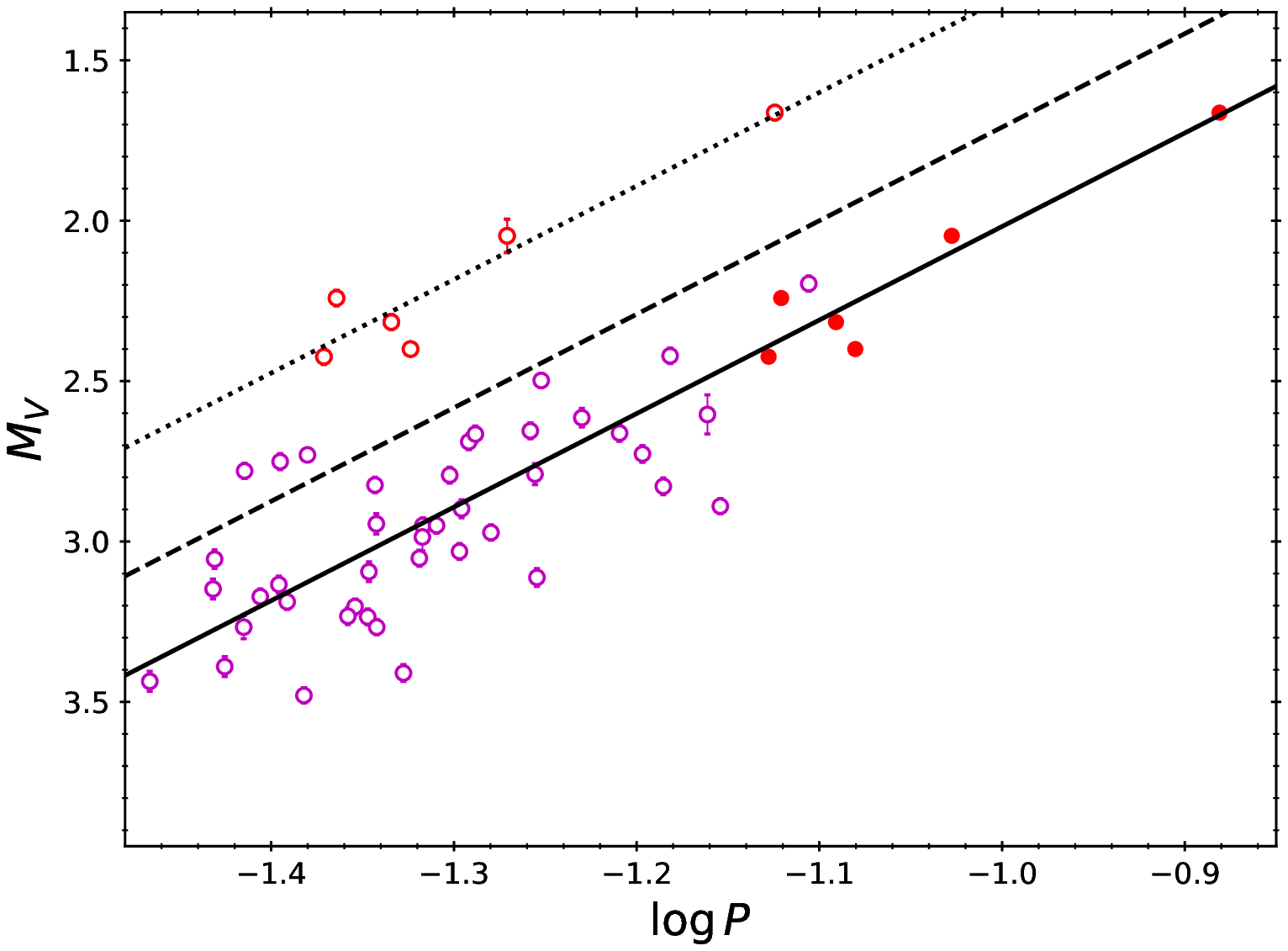}{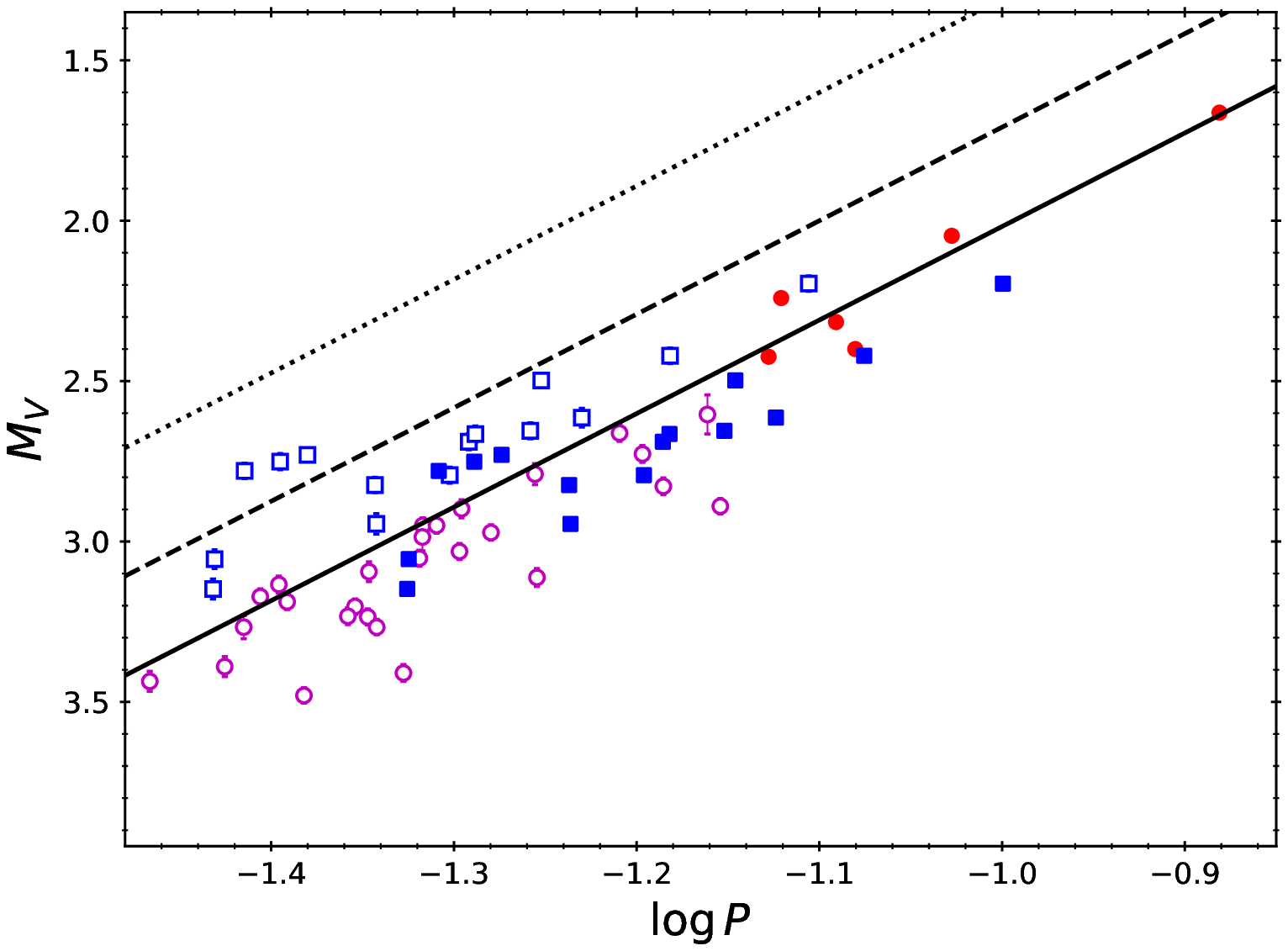}
  \caption{{\bf Left Panel:} The transformed $V$-band mean magnitudes for the 48 SXP stars, shown as open circles, listed in Table \ref{tab1} as a function of pulsation period $P$. The solid, dashed and dotted lines are the F, 1O, and 2O PL relation based on the $V$-band PL relation published in \citet[][see text for more details]{af2011}. SXP stars labeled as open red circles are assumed to pulsate in the second-overtone mode. After fundamentalizing their periods, their new locations on the PL plane are marked as filled red circles. {\bf Right Panel:} Same as the left panel, but for the 15 SXP stars identified as pulsating in the first-overtone mode, as described in the text. Open and filled blue squares are their locations on the PL plane before and after fundamentalizing their periods, respectively.}
  \label{fig_mode}
\end{figure*}

Mean magnitudes for SXP stars listed in Table \ref{tab1} were converted to absolute magnitudes by adopting the homogeneous distances ($D$) to their host globular clusters \citep{baumgardt2021}, together with reddening values $E$ returned from the {\tt Bayerstar2019} 3D reddening map \citep{green2019}\footnote{See \url{http://argonaut.skymaps.info/usage}}.

\subsection{Modes Identification with $V$-Band PL Relation}\label{sec4.1}

Since all of the SXP stars listed in Table \ref{tab1} have mean magnitudes in the $gr$-band, we converted these mean magnitudes to the $V$-band via the transformation provided in \citet{tonry2012}. Figure \ref{fig_plv} presents the extinction-corrected PL relation in the $V$-band, overlaid with several $V$-band PL relation for fundamental-mode (F) pulsators taken from literature (solid lines in Figure \ref{fig_plv}). The dashed and dotted lines in Figure \ref{fig_plv} are the PL relations for the first-overtone (1O) and second-overtone (2O) pulsators with the same PL relations as F pulsators, but shifted with a period ratio of $1O/F = 0.783$ and $2O/F = 0.571$ \citep[see][and reference therein]{af2011}, respectively. The PL relations from \citet{cohen2012} and \citet{martinazzi2015b} represent the two ``extremes'' among the available $V$-band PL relations, and the \citet{af2011} PL relation lies near the middle of them.

\begin{figure}
  \epsscale{1.05}
  \plotone{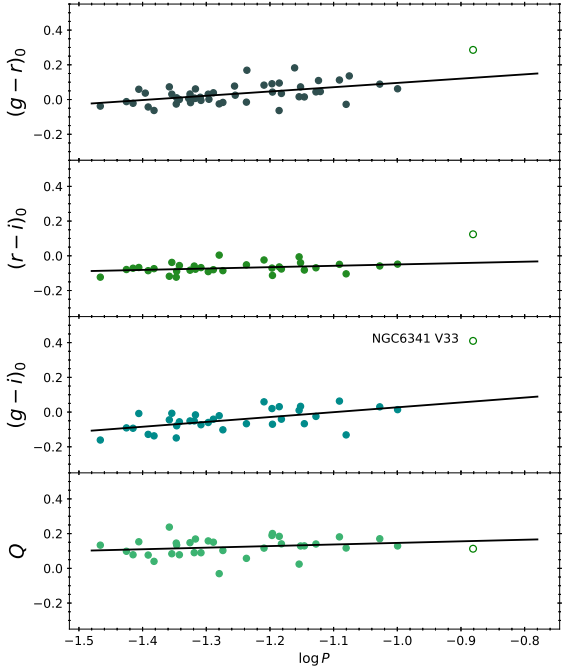}
  \caption{Extinction-corrected PC relations and the extinction-free PQ relations in the $gri$-band. Note that the periods for 1O and 2O SXP have been fundamentalized. The open circle marks the outlier NGC6341 V33 in the $(g-i)$ PC relation.}
  \label{fig_pcq}
\end{figure}

The six SXP stars located along the 2O PL relation from \citet{af2011} are clearly pulsating in the second-overtone mode. After fundamentalized their periods, they do lie along the fundamental-mode PL relation as shown in the left panel of Figure \ref{fig_mode}. Hence, these six SXP stars are classified as 2O SXP stars. The selection of 1O SXP stars, however, are nontrivial as their locations on the PL plane could be overlapped with F SXP stars. Therefore, we first calculated the residuals from the 1O PL relation for each SXP stars (after excluding the six 2O SXP stars), and rank these residuals in ascending order. We then fundamentalized the periods one-by-one for SXP stars in this list, and fit a PL relation with the rest of SXP stars. This procedure was terminated when the dispersion of the fitted PL relation reached the minimum, and we identified 15 SXP stars pulsating in the first-overtone mode. These 1O SXP stars are marked with squares in the right panel of Figure \ref{fig_mode}. The rest of the SXP stars were then classified as fundamental-mode SXP stars. The identified pulsation modes for our final sample of SXP stars are given in Table \ref{tab1}.

\begin{figure}
  \epsscale{1.05}
  \plotone{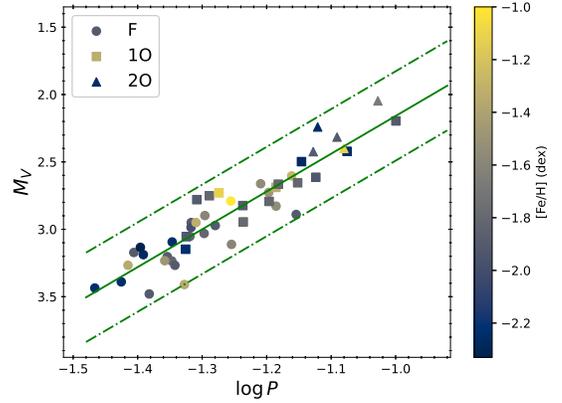}
  \caption{Fitted $V$-band PL relation, shown as the solid line, to the F (circles), 1O (squares), and 2O (triangles) SXP stars in the sample. The dotted-dashed lines represent the $\pm 2.5\sigma$ boundaries, where $\sigma$ is the dispersion of the fitted PL relation. The data points are color-coded with the metallicities of the host globular clusters ($\mathrm{[Fe/H]}$), adopted from the GOTHAM survey \citep[GlObular clusTer Homogeneous Abundances Measurements,][]{dias2015, dias2016a, dias2016b, vasquez2018}, where the metallicity ranged from $-1.00$~dex (NGC~6171) to $-2.33$~dex (NGC~7099).}
  \label{fig_vall}
\end{figure}

\subsection{The Multi-Band Relations}\label{sec4.2}

After fundamentalized the periods for 1O and 2O SXP stars, we first examined the period-color (PC) and period-$Q$-index (PQ) relations in the $gri$-band for these SXP stars, where $Q=(g-r)-1.395(r-i)$ is reddening-free by construction \citep[see][and reference therein]{ngeow2022b}, as presented in Figure \ref{fig_pcq}. NGC6341 V33 appears to be an outlier in the PC relations, especially the $(g-i)$ PC relation. Note that this star is $\sim1$~mag brighter in $r$-band than V41 in the same cluster (see Table \ref{tab1}), suggesting ZTF photometry for V33 could be affected by blending. Hence, we further excluded NGC6341 V33 when fitting for various relations. For the remaining SXP stars, the fitted PC relations are:

\begin{eqnarray}
  (g-r) & = & 0.247[\pm0.064] \log P + 0.342 [\pm0.081], \\
  (r-i) & = & 0.079[\pm0.042] \log P + 0.029 [\pm0.053], \\
  (g-i) & = & 0.280[\pm0.074] \log P + 0.308 [\pm0.093],
\end{eqnarray}

\noindent with a dispersion ($\sigma$) of $0.049$, $0.029$, and $0.050$, respectively. Similarly, the derived PQ relation is
\begin{eqnarray}
  Q & = & 0.091[\pm0.080] \log P + 0.238 [\pm0.101],\sigma=0.054. 
\end{eqnarray}

We then fit a PL relation to the transformed $V$-band mean magnitudes and obtained the following $V$-band PL relation:

\begin{eqnarray}
  M_V & = & -2.804[\pm0.035] \log P - 0.645 [\pm0.044], \\
  \sigma_V & = & 0.133. \nonumber
\end{eqnarray}

\noindent The fitted fundamental-mode $V$-band PL relation, shown in Figure \ref{fig_vall}, is similar to the theoretical $V$-band PL relation presented in \citet[][$-2.847\log P - 0.45$ at $Z=0.0001$]{fiorentino2015}. Since there is no clear trend of the metallicity on the $V$-band PL relation, as displayed in Figure \ref{fig_vall}, as well as a rather small sample size with 47 SXP stars, we do not include a metallicity term when fitting the PL relation. 

In a similar manner, we derived the following $gri$-band PL relations:

\begin{eqnarray}
  M_g & = & -2.719[\pm0.031] \log P - 0.529 [\pm0.039], \\
  \sigma_g & = & 0.135, \nonumber \\
  M_r & = & -2.907[\pm0.031] \log P - 0.792 [\pm0.039], \\
  \sigma_r & = & 0.135, \nonumber \\
  M_i & = & -2.917[\pm0.035] \log P - 0.737 [\pm0.043], \\
  \sigma_i & = & 0.129. \nonumber
\end{eqnarray}

\noindent For the PW relations, following our previous work \citep[see][and reference therein]{ngeow2022b}, we obtained the three PW relations as given below:

\begin{eqnarray}
  W_g^{gr} & = & -3.458[\pm0.031] \log P - 1.561 [\pm0.038], \\
  W_r^{ri} & = & -3.092[\pm0.034] \log P - 0.743 [\pm0.043], \\
  W_g^{gi} & = & -3.195[\pm0.034] \log P - 1.024 [\pm0.043]. 
\end{eqnarray}

\begin{figure*}
  \epsscale{1.05}
  \plottwo{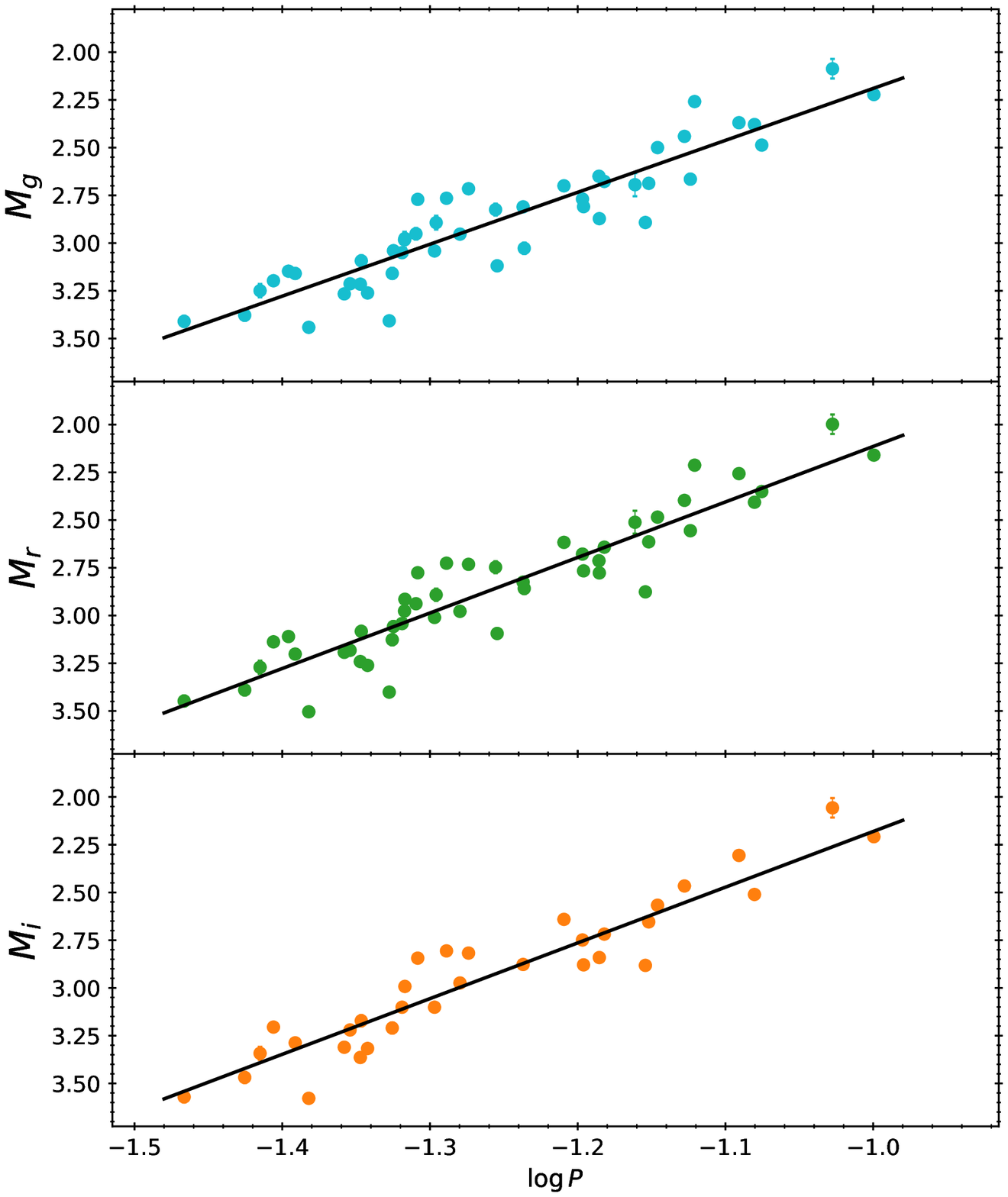}{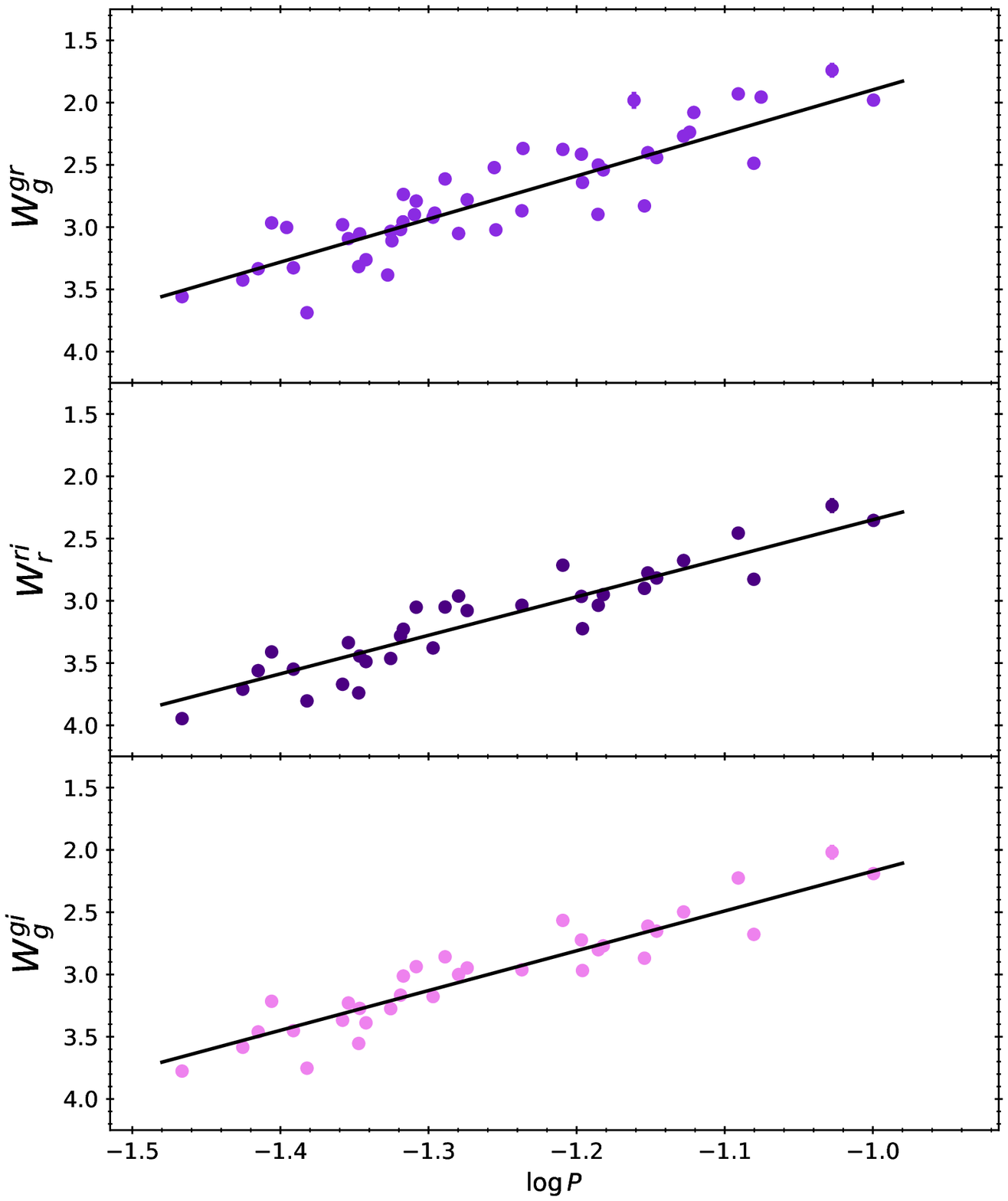}
  \caption{Extinction-corrected PL relations (left panel) and the extinction-free PW relations (right panel) in the $gri$-band for our sample of SXP stars in the globular clusters. The solid lines are the fitted relations as given in equation (6) to (11).}
  \label{fig_plw}
\end{figure*}

\noindent Dispersions of these PW relations, however, are larger than the PL relations, with $\sigma=0.216$, $0.163$, and $0.161$, respectively. These $gri$-band PL and PW relations are presented in Figure \ref{fig_plw}.

\section{Example of Application: The Only SXP Star in Crater II} \label{sec5}

\citet{vivas2020} reported the detection of one SXP star, with an ID of V97, in dwarf galaxy Crater II using $gi$-band time-series DECam observations. This SXP star has a rather long pulsation period, $P=0.23461$~days, much longer than the SXP stars listed in Table \ref{tab1}. Assuming V97 is pulsating in fundamental-mode, then the predicted absolute magnitudes at this period, based on equation (6) and (8), are $M_g(V97)=1.18\pm0.18$~mag and $M_i(V97)=0.79\pm0.19$~mag, respectively. Since the photometry presented in \citet{vivas2020} were calibrated to the SDSS photometric system, we transformed the mean $gi$-band magnitudes for V97, after correcting for extinction using the {\tt Bayerstar2019} 3D reddening map \citep{green2019}, to the PAN-STARRS1 system based on the transformations given in \citet{tonry2012}. These transformations, however, required the $(g-r)$ color for the target star, which is absent from \citet{vivas2020} observations. Hence, we estimated the $(g-r)$ color using equation (1), which gives a predicted color of $0.19\pm0.06$~mag for V97. The resulted $gi$-band distance moduli are $\mu_g=20.02\pm0.18$~mag and $\mu_i=20.02\pm0.18$~mag, respectively. Similarly, the predicted absolute $W_g^{gi}$ Wesenheit magnitude is $M_W(V97)=0.99\pm0.22$~mag, which give a (extinction-free) distance modulus of $\mu_W=20.03\pm0.23$~mag.

Clearly, these distance moduli are smaller than $\mu\sim 20.33$~mag found by \citet{vivas2020} based on RR Lyrae. \citet{vivas2020} suspected that V97 could be pulsating in the first-overtone mode. If we fundamentalized the period of V97 and repeated the above calculations, the resulted distance moduli became $\mu_g=20.31\pm0.19$~mag, $\mu_i=20.33\pm0.19$~mag, and $\mu_W=20.37\pm0.24$~mag. These distance moduli are consistent with the values found by \citet{vivas2020} and \citet{ngeow2022a}, supporting the assumption that V97 is indeed pulsating in the first-overtone mode.

\section{Conclusions} \label{sec6}

Based on homogeneous distances to globular clusters, we calibrated the PL and PW relations for SXP stars in the $gri$-band using the light curves data obtained from ZTF. In addition, we have also updated the $V$-band PL relation after transforming the $gr$-band mean magnitudes to the $V$-band for our sample of SXP stars, as well as deriving the $gri$-band PC relations and PQ relation for the first time. We tested our derived relations on the only SXP star discovered in dwarf galaxy Crater II. Assuming this SXP star is pulsating in first-overtone mode, the derived distance moduli to Crater II using our PL and PW relations are consistent with value given in the literature.

In coming years when LSST is in its routine operations, it is expected to detect more SXP variables in Crater II. According to our derived PL relations and adopting the distance modulus of Crater II to be $\sim 20.33$~mag, the shortest period for SXP stars at $\log P\sim -1.5$ (or $\sim 0.032$~days) would have a mean magnitude of $\sim23.9$~mag in the $g$-band, which is brighter than the detection limit of $25.0$~mag from a single epoch LSST observations \citep{lsst2019}. Hence, it is possible that LSST could detect additional SXP stars in Crater II, as well as finding new SXP stars in other dwarf galaxies with similar distances to Crater II. Our empirical $gri$-band PL and PW relations for SXP variables will improve the distance determination using these population II short period pulsating stars, and provide an independent cross-check to other distance indicators such as RR Lyrae and Type II Cepheids.

\acknowledgments

We thank the useful discussions and comments from an anonymous referee that improved the manuscript. We are thankful for funding from the National Science and Technology Council (Taiwan) under the contracts 107-2119-M-008-014-MY2, 107-2119-M-008-012, 108-2628-M-007-005-RSP and 109-2112-M-008-014-MY3.

Based on observations obtained with the Samuel Oschin Telescope 48-inch Telescope at the Palomar Observatory as part of the Zwicky Transient Facility project. ZTF is supported by the National Science Foundation under Grants No. AST-1440341 and AST-2034437 and a collaboration including current partners Caltech, IPAC, the Weizmann Institute of Science, the Oskar Klein Center at Stockholm University, the University of Maryland, Deutsches Elektronen-Synchrotron and Humboldt University, the TANGO Consortium of Taiwan, the University of Wisconsin at Milwaukee, Trinity College Dublin, Lawrence Livermore National Laboratories, IN2P3, University of Warwick, Ruhr University Bochum, Northwestern University and former partners the University of Washington, Los Alamos National Laboratories, and Lawrence Berkeley National Laboratories. Operations are conducted by COO, IPAC, and UW.

This research has made use of the SIMBAD database and the VizieR catalogue access tool, operated at CDS, Strasbourg, France. This research made use of Astropy,\footnote{\url{http://www.astropy.org}} a community-developed core Python package for Astronomy \citep{astropy2013, astropy2018, astropy2022}.

\facility{PO:1.2m}

\software{{\tt astropy} \citep{astropy2013,astropy2018,astropy2022}, {\tt dustmaps} \citep{green2018}, {\tt gatspy} \citep{vdp2015}, {\tt Matplotlib} \citep{hunter2007},  {\tt NumPy} \citep{harris2020}, {\tt SciPy} \citep{virtanen2020}, {\tt statsmodels} \citep{seabold2010}.}

%\newpage

% ===============================================
%               REFERENCE
% ===============================================

\end{document}